\documentclass[aps,prb,twocolumn,showpacs]{revtex4-1}
\pdfoutput=1
\usepackage{graphicx,epsf}
\usepackage{amsmath}
\usepackage{amssymb}
\usepackage{color}

\newcommand{\bR}{{\bf R}}
\newcommand{\bQ}{{\bf Q}}
\newcommand{\bq}{{\bf q}}
\newcommand{\bp}{{\bf p}}
\newcommand{\bk}{{\bf k}}
\newcommand{\br}{{\bf r}}
\newcommand{\bqz}{{\bf Q}}

\begin{document}

\title{Finite--momentum condensate of magnetic excitons
         in a bilayer quantum Hall system}

\author{R. L. Doretto$^1$}
\author{C. Morais Smith$^2$}
\author{A. O. Caldeira$^3$}

\affiliation{$^1$Instituto de F\'isica Te\'orica, Universidade
                 Estadual Paulista,
                 01140-070 S\~ao Paulo, SP, Brazil \\
             $^2$Institute for Theoretical Physics, Utrecht
                 University, 3584 CE Utrecht, The
                 Netherlands \\
             $^3$Instituto de F\'isica Gleb Wataghin, Universidade
                 Estadual de Campinas, 13083-970 Campinas, SP, Brazil
             }
\date{\today}

\begin{abstract}

We study the bilayer quantum Hall system at total filling factor
$\nu_T = 1$ within a bosonization formalism which allows us to approximately
treat the magnetic exciton as a boson. We show that in the region
where the distance between the two layers is comparable to the magnetic length,
the ground state of the system can be seen as a finite--momentum
condensate of magnetic excitons provided that the excitation spectrum
is gapped. We analyze the stability of such a phase within the
Bogoliubov approximation firstly assuming that only one momentum 
$\bQ$ is macroscopically occupied  and later we consider the same
situation for two modes $\pm\bQ$.       
We find strong evidences that a first--order quantum phase transition
at small interlayer separation takes place  from a zero--momentum
condensate phase, which corresponds to Halperin 111 state, to a
finite--momentum condensate of magnetic excitons.

\end{abstract}

\maketitle

%
%
%
%

\maketitle

\section{Introduction}
\label{sec:intro}

A bilayer quantum Hall system (QHS) consists of two two-dimensional
electron gases (layers) separated by a small distance $d$ under an
uniform  magnetic field ${\bf B}$ perpendicular to the layers. Among
the several possible configurations, we consider the one where each
layer has filling factor $\nu = n\phi_0/B = 1/2$, such that the total filling
factor $\nu_T = 1/2 + 1/2=1$. Here, $n$ is the electronic density of
each layer and $\phi_0 = hc/e$, the magnetic flux
quantum.\cite{book-qhe,eisenstein04}

The system is characterized by two parameters: the ratios $d/\ell$ and
$\Delta_{\rm SAS}/E_c$. Here, $\ell = \sqrt{\hbar c/eB}$ is the
magnetic length, the characteristic length scale of QHSs, $\Delta_{\rm
SAS}$ is the electron interlayer tunneling energy, and $E_c =
e^2/\epsilon\ell$ is the characteristic Coulomb energy with $\epsilon$ being
the dielectric constant of the host semiconductor. Although
$\Delta_{\rm SAS}$ and the distance $d$ are fixed for a given sample,
the ratio $d/\ell$ can be modified by changing  the magnetic field
${\bf B}$ and then adjusting the electronic density in each layer in such
a way that the configuration $\nu_T = 1/2 + 1/2 = 1$ is restored.
Interestingly, a series of measurements\cite{exp-bilayer,spielman00,spielman01} 
has shown that for $d < d_c \approx 1.8\, \ell$, the bilayer QHS
behaves as a single--layer QHS at $\nu=1$,
while  for $d > d_c$, as two independent two-dimensional
electron gases at $\nu=1/2$. In spite of the fact that the
experimental data indicate a continuous transition between these two
situations, the so--called incompressible--compressible quantum phase
transition, from the theoretical point of view it is not clear whether 
the system undergoes a second--order quantum phase transition\cite{moller09} or a
first--order one smeared out by disorder.\cite{schliemann01,ye07}

The ground state of the bilayer QHS at $\nu_T=1$ is well understood in
two limiting cases: for small $d/\ell$, it can be described by the
(incompressible) Halperin $111$ wave function,\cite{halperin83} while
in the very large $d/\ell$ region, by two independent
(compressible) composite fermion Fermi
liquids.\cite{halperin93,heinonen}  Interestingly, the Halperin $111$ state
can be seen as a Bose-Einstein condensate (BEC) of magnetic excitons,
where the electron and the hole are in different
layers.\cite{fertig89} This analogy motivated us to employ the bosonization
scheme\cite{doretto05} to study the bilayer QHS at $\nu_T =1$. Our
main finding in this first study\cite{doretto06} was 
that a {\it zero}--momentum  BEC of
magnetic excitons is stable only for $d \le 0.4\,\ell$ (zero interlayer
tunneling case). Such a result is in quite good agreement with the
exact diagonalization calculations on finite size systems, which
show that the overlap between the exact ground state
and the $111$ state is close to unit only for $d \lesssim
0.5\,\ell$.\cite{simon03,moller08}

\begin{figure*}[t]
\centerline{\includegraphics[width=4.5cm]{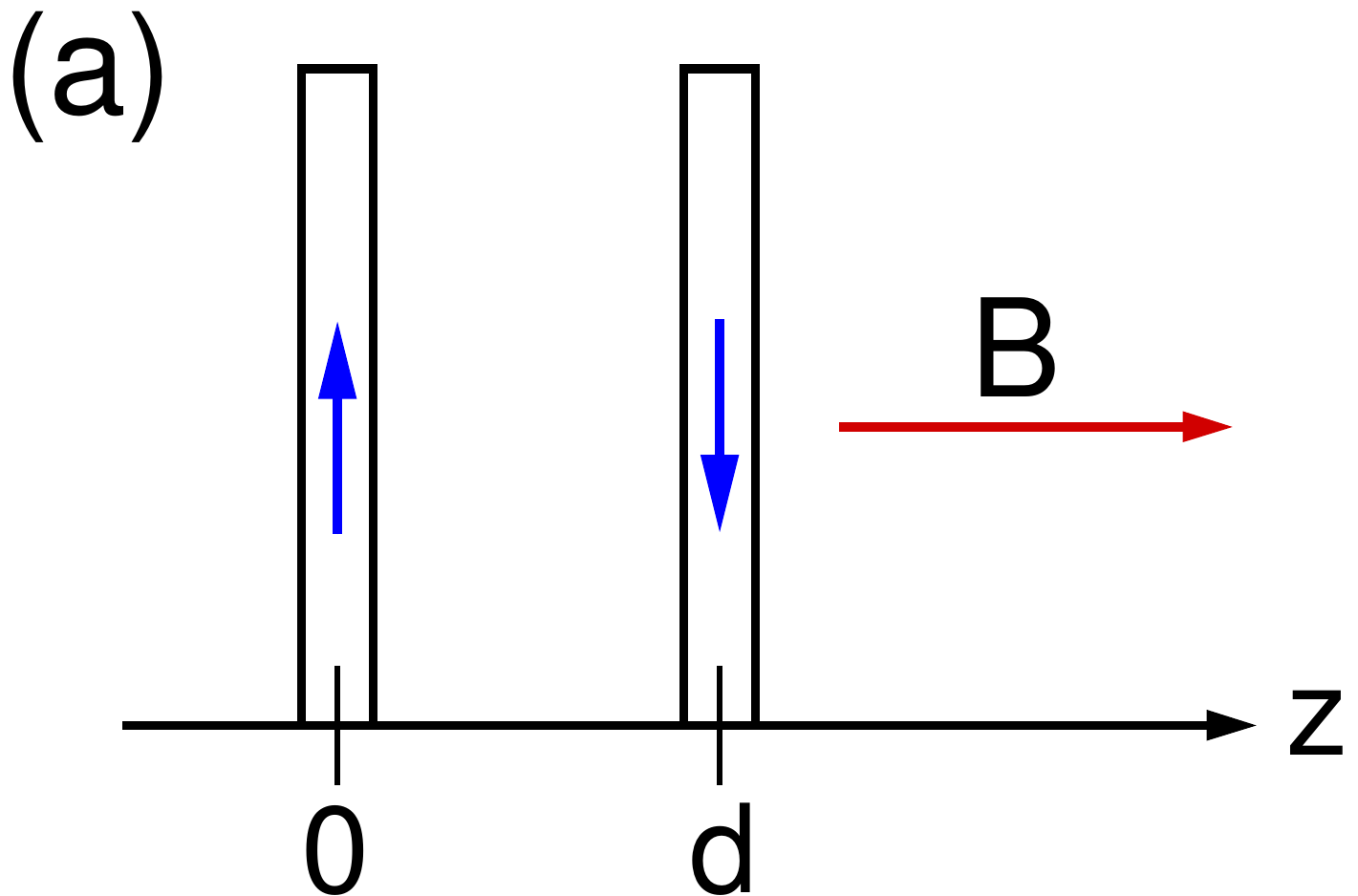}
            \hskip0.7cm
            \includegraphics[width=12.2cm]{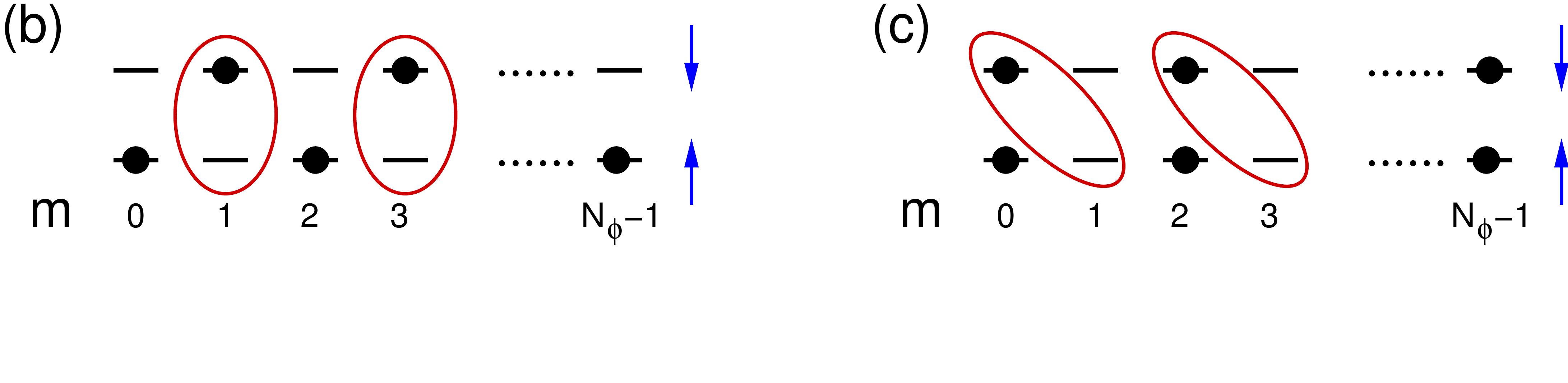}
           }
\caption{(Color online) Schematic representations: (a) Bilayer
  QHS. ${\bf B}$
  is the magnetic field and $d$ the distance between the two layers.
  (b) Zero--momentum BEC of magnetic excitons and (c) finite--momentum BEC
  of magnetic excitons with $|\ell\bqz| = 1$. Only the lowest Landau
  levels corresponding to the $\uparrow$
  and $\downarrow$ layers are shown. $m$ is the guiding center quantum
  number which labels the degeneracy of each Landau level.}
\label{fig:bilayer}
\end{figure*}

Although much theoretical
work\cite{ye07,simon03,moller08,cote92,kim01,nomura02,
schliemann03,park04,moller09,milica07}
has been devoted to the intermediate region, $d \sim \ell$, so
far there is no consensus about the nature of the ground state. For
instance, a (pseudospin) density wave,\cite{ye07} mixed Fermi--Bose
trial wave functions,\cite{simon03} and a (pseudospin) spiral
state\cite{park04} have been proposed as possible candidates. 
A proper description of the ground state in the
intermediate $d/\ell$ region is important since it will help us to
determine the nature of the incompressible--compressible phase
transition.

In this paper, we revisit the bilayer QHS within the bosonization
formalism\cite{doretto05,doretto06} focusing on the intermediate
$d/\ell$ region. We propose that within this bosonic scheme the ground
state of the system can be seen as a {\it finite}--momentum BEC of
magnetic excitons. We show that this is indeed a possible phase of the
effective boson model that we have derived in Ref.~\onlinecite{doretto06},
provided that the (neutral) quasiparticle excitation spectrum is gapped.
Our results also indicate that the instability of the zero--momentum
BEC of magnetic excitons at $d = 0.4\,\ell$ reported in
Ref.~\onlinecite{doretto06} indeed corresponds to a first--order
quantum phase transition from a zero--momentum BEC of magnetic
excitons to a finite--momentum one.

Our paper is organized as follows: In Sec.~\ref{sec:model}, we
introduce an interacting fermion model to describe the bilayer QHS,
summarize the bosonization method,\cite{doretto05} and recall the
main steps to derive the
effective boson model from the original fermionic one. We also comment
on the motivation for considering a finite--momentum BEC of magnetic
excitons as the ground state of the bilayer. Sec.~\ref{sec:onemode} is
devoted to the analysis of the effective boson model within the
Bogoliubov approximation assuming that the ground state is given by a
finite--momentum BEC of magnetic excitons where the momentum $\bqz =
Q\hat{x}$ is macroscopically occupied. The ground state energy
and the (neutral) quasiparticle excitation spectrum are
calculated. Here, evidences that a first--order quantum phase
transition takes place at small $d/\ell$ are found. In
Sec.~\ref{sec:twomodes}, we perform a similar analysis but now
considering that two modes, $\pm\bqz$ with $\bqz =
Q\hat{x} \not= 0$, are macroscopically occupied. We show that $|l\bqz_0|$,
the magnitude of the momentum associated with the lowest energy configuration,
increases with $d/\ell$. Some additional features of a BEC of magnetic
excitons are shown in Sec.~\ref{sec:properties}. 
In Sec.~\ref{sec:discussion}, we compare our
results with previous ones and comment on their consequences for the
bilayer QHS at $\nu_T =1$. A short summary with the main results
closes the paper. The fact that density fluctuations can account for
the definition of boson operators for the bilayer QHS, comparison with
alternative bosonic schemes used to describe the bilayer, and some
details of the calculations can be found in the Appendices.

\section{Model}
\label{sec:model}

Let us consider a two layer system composed of $N$ electrons moving
in the $(x,y,z=0)$ plane and $N$ in the $(x,y,z=d)$ plane under an
external magnetic field ${\bf B} = B\hat{z}$,
Fig.~\ref{fig:bilayer}(a), at zero temperature. We introduce a
pseudospin index $\alpha = \uparrow , \downarrow$ in order to label
each layer. We also assume 
that the ${\bf B}$ field is strong enough such that the electrons are
fully spin polarized (frozen electronic spin degree of freedom) and
that the Hilbert space of each layer is restricted to the
corresponding lowest Landau level. The configuration $\nu_T
=\nu_\uparrow + \nu_\downarrow = 1/2 + 1/2 = 1$ is realized by setting
the degeneracy of each Landau level $N_\Phi = 2N$.

The Hamiltonian of the system has only two terms (since all electrons
are restricted to the lowest Landau level, the kinetic energy is a
constant and can be neglected): 
\begin{equation}
 H = H_T + H_I.
\label{ham}
\end{equation}
Here, $H_T$ describes the electron tunneling between the two layers,
\begin{equation}
  H_T = -\frac{1}{2}\Delta_{\rm SAS}\sum_m
         c^\dagger_{m\,\uparrow}c_{m\,\downarrow}  + {\rm H.c.},
\label{ham-t}
\end{equation}
and $H_I$ is the Coulomb interaction term (we set the system area
$\mathcal{A}=1$),
\begin{equation}
  H_I = \frac{1}{2}\sum_{\bk\not= 0}\sum_{\alpha\beta = \uparrow , \downarrow}
            v_{\alpha\beta}(k)\rho_\alpha(\bk)\rho_\beta(-\bk)
\label{ham-i}
\end{equation}
with $k = |\bk|$. $\Delta_{\rm SAS}$ is the electron interlayer
tunneling energy, $c^\dagger_{m\alpha}$ creates an electron with
guiding center $m$ in the lowest Landau level of the $\alpha$ layer,
Fig.~\ref{fig:bilayer}(b), and $\rho_\alpha(\bk)$ is the Fourier
transform of the $\alpha$--electron density operator projected into
the lowest Landau level, i.e.,\cite{doretto05}
\begin{equation}
  \rho_{\alpha}(\bk)=e^{-(\ell k)^{2}/4}\sum_{m,m'}G_{m,m'}(\ell\bk)
                     c_{m\alpha}^{\dagger}c_{m'\alpha}.
\label{density-op}
\end{equation}
The function $G_{m,m'}(x)$ is defined in the Appendix C of
Ref.~\onlinecite{doretto05}. Finally, 
\begin{eqnarray}
v_{\uparrow\uparrow}(k) &=& v_{\downarrow\downarrow}(k) = v_A(k)
    = \frac{2\pi e^2}{\epsilon k}e^{-(\ell k)^2/2},
\nonumber \\
v_{\uparrow\downarrow}(k) &=& v_{\downarrow\uparrow}(k) = v_E(k)
    = \frac{2\pi e^2}{\epsilon k}e^{-(\ell k)^2/2}e^{-kd}
\end{eqnarray}
are, respectively, the Fourier transforms of the intralayer, 
$v_A(r) =  e^2/\epsilon r$,  and interlayer,
$v_E(r) = e^2/\epsilon\sqrt{r^2 + d^2}$, electron--electron
interaction potentials with $r = |\br|$. 

On can show that Eq.~\eqref{ham-t} can
be written in terms of the $x$--component of the pseudospin density
operator, i.e.,
\begin{equation}
 H_T = -\Delta_{\rm SAS}S_x(\bk = 0),
\end{equation}
while Eq.~\eqref{ham-i}, in terms of the total electron
density operator $\rho(\bk) = \rho_\uparrow(\bk) +
\rho_\downarrow(\bk)$ and the $z$-component of the pseudospin density
operator $S_z(\bk) = [\rho_\uparrow(\bk) - \rho_\downarrow(\bk)]/2$,
namely
\begin{equation}
  H_I = \frac{1}{2}\sum_{\bk\not= 0} v_0(k)\rho(\bk)\rho(-\bk)
         + 2\sum_{\bk\not= 0}v_z(k)S_z(\bk)S_z(-\bk),
\label{ham-i2}
\end{equation}
with
\begin{equation}
   v_{0/z}(k) = \frac{1}{2}[v_A(k) \pm v_E(k)]
              = \frac{\pi e^2}{\epsilon k}e^{-(\ell k)^2/2}
                \left(1 \pm e^{-kd}\right).
\end{equation}
In the following, we focus on the zero tunneling case, i.e., we set
$\Delta_{\rm SAS} = 0$ which yields $H = H_I$.

\subsection{Bosonization formalism}
\label{sec:boso}

We study the interacting fermion model \eqref{ham-i2} within the
bosonization formalism\cite{doretto05} that was recently developed by
two of us among others. Although such a scheme was originally proposed for the {\it
single}--layer QHS at $\nu=1$, it is possible to show that it also
holds for the bilayer QHS at $\nu_T =1$, see Appendix \ref{ap:boso}.
We now briefly summarize the bosonization method and refer the reader
to Ref.~\onlinecite{doretto05} for more details. In Appendix
\ref{ap:alternative}, we briefly comment on some alternative bosonic
descriptions employed to study the bilayer QHS.

\begin{figure}[t]
\centerline{\includegraphics[width=6.1cm]{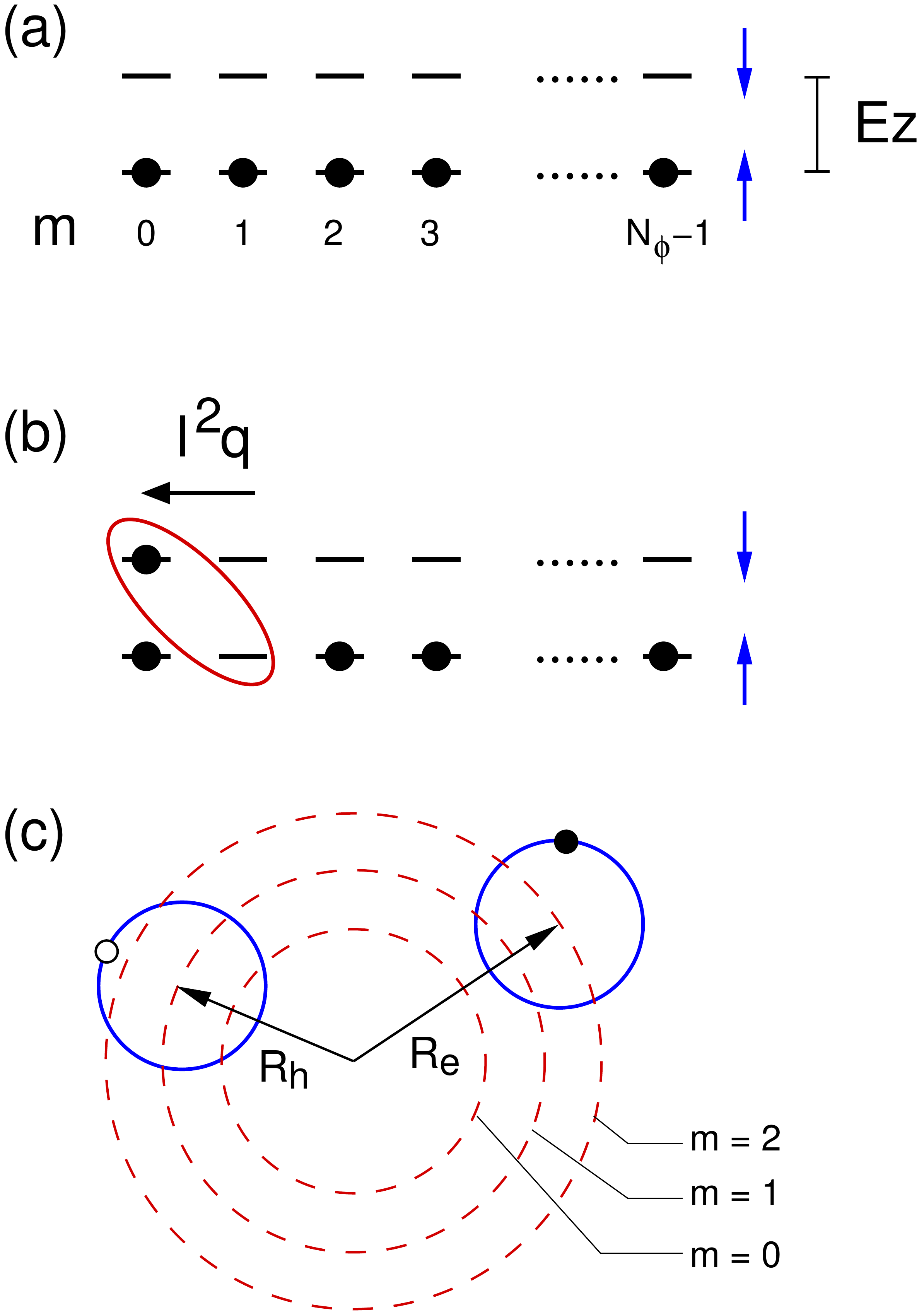}}
\caption{(Color online) Schematic representations: (a) Quantum Hall
  ferromagnet $|{\rm FM}\rangle$, the ground state of the
  single--layer QHS at $\nu = 1$, and (b) an electron--hole pair
  excitation (magnetic exciton) with momentum $|l\bq| = 1$ above
  $|{\rm FM}\rangle$. Only the spin up and spin down lowest Landau
  levels are shown. $E_z$ is the Zeeman energy and $m$ is the guiding
  center quantum number. (c) Semiclassical representation of an
  electron--hole pair in the symmetric gauge. Particles move along
  cyclotron orbits (solid blue circles) whose center are at one of the
  possible guiding centers (dashed red circles). The vectors $\bR_e$
  and $\bR_h$ correspond to the guiding center position of the
  electron (solid circle) and the hole (open circle), respectively.}
\label{fig:qhfm}
\end{figure}

Let us consider the single--layer QHS at $\nu=1$. We restrict the
Hilbert space to the lowest Landau level and explicitly take into
account the electronic spin. The ground state of the system, the
so--called quantum Hall ferromagnet $|{\rm FM}\rangle$, is illustrated
in Fig.~\ref{fig:qhfm}(a). It is possible to show that the
neutral elementary excitations above this state, electron--hole pairs also
known as magnetic excitons, Fig.~\ref{fig:qhfm}(b),
can be approximately treated as bosons.
More precisely, we can define the following bosonic operators:
\begin{eqnarray}
   b_\bq &=& N^{-1/2}_\Phi e^{-(\ell q)^2/4}\sum_{m,m'}
   G_{m,m'}(-\ell\bq)c^\dagger_{m\,\uparrow}c_{m'\,\downarrow},
\nonumber \\
   b^\dagger_\bq &=& N^{-1/2}_\Phi e^{-(\ell q)^2/4}\sum_{m,m'}
   G_{m,m'}(\ell\bq)c^\dagger_{m\,\downarrow}c_{m'\,\uparrow},
\label{boson-op}
\end{eqnarray}
where $c^\dagger_{m\,\sigma}$ ($c_{m\,\sigma}$) is a creation
(annihilation) operator for an electron in the lowest Landau level with
guiding center $m$ and spin $\sigma$. The boson operators
\eqref{boson-op} obey the canonical commutation relations
$[b^\dagger_\bq, b^\dagger_\bk] = [b_\bq, b_\bk] = 0$ and $[b_\bq,
b^\dagger_\bk] = \delta_{\bq, \bk}$ once some conditions are
fulfilled. The state $b^\dagger_\bq|{\rm FM}\rangle$ corresponds to a
magnetic exciton with momentum $\bq$, Fig.~\ref{fig:qhfm}(b). Within this
framework, the electron density operator and the $z$--component of the
spin density operator read
\begin{eqnarray}
\rho(\bk) & = & \delta_{k,0}N_\Phi +2i\sum_\bq
                \sin(\bk\wedge\bq/2)b_{\bk+\bq}^\dagger b_\bq,
 \label{rho-boso} \\
S_z(\bk) & = & \frac{1}{2}\delta_{k,0}N_\Phi -\sum_\bq
               \cos(\bk\wedge\bq/2)b_{\bk+\bq}^\dagger b_\bq
\label{sz-boso}
\end{eqnarray}
with $\bk\wedge\bq \equiv \ell^2\hat{z}\cdot(\bk\times\bq)$.

It is easy to see that, in principle, the bosonization scheme outlined
above can be employed to study the bilayer QHS at $\nu_T = 1$, once
the pseudospin $\alpha$ is identified with the electronic spin quantum
number $\sigma$ of the single--layer QHS at $\nu=1$ [compare
Figs~\ref{fig:bilayer}(b) and \ref{fig:qhfm}(a) and recall that we
consider that the electrons are completely spin polarized in the
bilayer QHS]. Since the bosons $b$
are defined with respect to a reference state, the quantum Hall
ferromagnet $|{\rm FM}\rangle$, the bilayer QHS at $\nu_T = 1/2 + 1/2
= 1$ corresponds to a system with $N_\Phi/2$ bosons, as illustrated in 
Fig.~\ref{fig:bilayer}(b).

\subsection{Effective boson model}

\begin{figure}[t]
\centerline{\includegraphics[width=7.1cm]{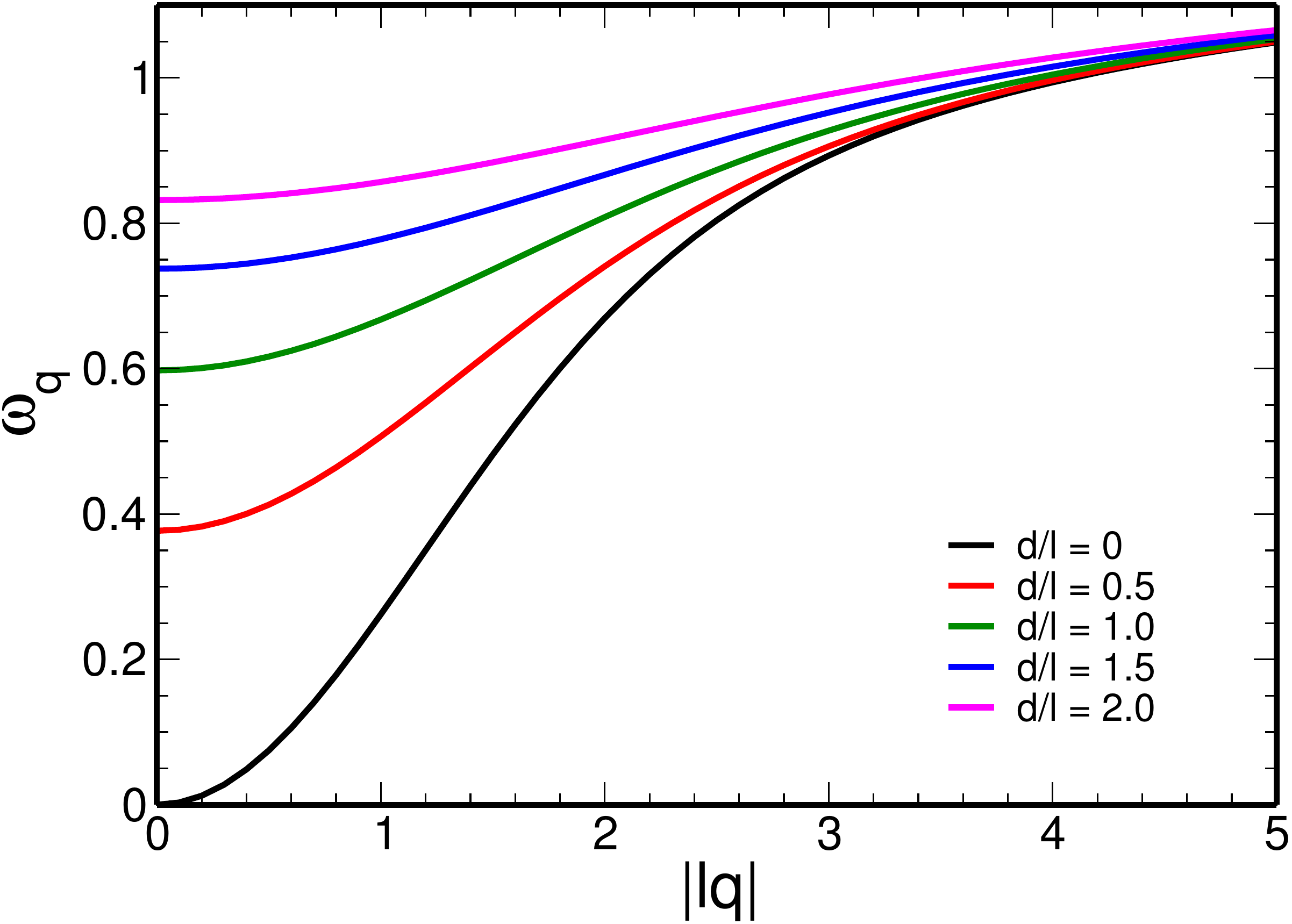}}
\caption{(Color online) Dispersion relation of the free bosons (in
  units of $e^2/\epsilon \ell$), Eq.~\eqref{energy-oneboson}, for $d/\ell =
  0$, $0.5$, $1$, $1.5$, and $2$ (from bottom to top at $\ell q = 0$).}
\label{fig:disp-bosons}
\end{figure}

Let us now follow the lines of Ref.~\onlinecite{doretto05} and map the
original interacting fermion model \eqref{ham-i2} into an effective
interacting boson model. Substituting Eqs.~\eqref{rho-boso} and
\eqref{sz-boso} into Eq.~\eqref{ham-i2} and normal ordering the
result, we arrive at
\begin{equation}
  H_B = \sum_\bq \omega_\bq b^\dagger_\bq b_\bq +
            \sum_{\bk\not= 0,\bp,\bq} v_\bk(\bp,\bq)
            b^\dagger_{\bk + \bp}b^\dagger_{\bq - \bk}b_\bq b_\bp.
\label{ham-boso}
\end{equation}
Here,\cite{note01}
\begin{equation}
\omega_\bq =  \frac{e^2}{\epsilon l}\left[\sqrt{\frac{\pi}{2}} -
                l\int_0^\infty dk\, e^{-kd}e^{-(kl)^2}J_0(kql^2)\right]
\label{energy-oneboson}
\end{equation}
is the dispersion relation of the free bosons (see Fig.~\ref{fig:disp-bosons}),
with $J_0(x)$ denoting the Bessel function of the first kind and
\begin{eqnarray}
  v_\bk(\bp,\bq) &=& 2v_0(k)\sin(\bk\wedge\bp/2)\sin(\bk\wedge\bq/2)
\nonumber \\
        &+&          2v_z(k)\cos(\bk\wedge\bp/2)\cos(\bk\wedge\bq/2)
\label{boson-potential}
\end{eqnarray}
is the boson--boson interaction potential. 
In the following, instead of $H_B$, we consider
\begin{equation}
 K = H_B - \mu\hat{N},
\label{ham-k}
\end{equation}
which explicitly includes the chemical potential $\mu$. Here, $\hat{N}
= \sum_\bq b^\dagger_\bq b_\bq$ is the number operator for bosons.

\subsection{Finite--momentum BEC of magnetic excitons}
\label{sec:ansatz}

In Ref.~\onlinecite{doretto06}, we analyzed the interacting boson model
\eqref{ham-boso} assuming that the bosons $b$ condense in their lowest
energy state, the $\bq = 0$ mode, and showed that such a state is stable
only for $d \le 0.4\,\ell$. The good agreement between our results and
exact diagonalization calculations, see the Introduction section,
tells us that not only is this zero--momentum BEC a good approximation
for Halperin $111$ state, but also that the bosonic
formalism\cite{doretto05} is indeed quite appropriate to describe the
bilayer QHS at $\nu_T=1$. Therefore, it might be possible to describe
the decrease of the correlations between the two layers as $d/\ell$
increases, i.e., the intermediate $d/\ell$ region, using solely the
bosonic degrees of freedom. In this case, what should be the form of
the ground state in terms of the bosons $b$ for $d \sim \ell$?

In order to construct the new ground state, we should recall some
properties of the magnetic excitons. As mentioned above, the state
$b_\bq^\dagger|{\rm FM}\rangle$ corresponds to a magnetic exciton with
momentum $\bq$, which is nothing but a suitable linear
combination of electron--hole pairs above the $|{\rm FM}\rangle$
state, see Eq.~\eqref{boson-op}. 
The momentum $\bq$ is canonically conjugate to the vector
$\bR_0 = (\bR_e + \bR_h)/2$,\cite{doretto05} where the vectors $\bR_e$
and $\bR_h$ denote respectively the position of the guiding
centers of the electron and the hole as illustrated in
Fig.~\ref{fig:qhfm}~(c). Interestingly, it is also possible to show that
[see Eq.~(2.16) in  Ref.~\onlinecite{kallin84}]
\begin{equation}
 \langle {\rm FM}| b_\bq (\bR_e - \bR_h) b^\dagger_\bq
                 |{\rm FM}\rangle = \ell^2\bq\times\hat{z},
\end{equation}
i.e., the (relative) distance between the guiding centers of the electron and
the hole which constitute the magnetic exciton is $\propto\, q$. Note that
this is an unusual relation between momentum and distance. Therefore,
a boson $b$ with $\bq = 0$ can be seen as an electron--hole pair 
both localized in the same guiding center, while for a boson $b$ with
$\bq \not= 0$, the electron and the hole are in different guiding
centers.

A zero--momentum BEC of magnetic excitons is then characterized by a
large number of (interlayer) electron--hole pairs where each electron
is very close (in the guiding center sense) to its partner hole as
depicted in Fig.~\ref{fig:bilayer}~(b). Since this is the smallest
distance between the electron and the hole, such a feature indicates
that the two layers are highly correlated, corroborating the relation
between the zero--momentum BEC and the $111$ state. Therefore, in order to
decrease the coupling between the two layers, we should, in principle,
consider a state constituted of a large number of electron--hole pairs where
now each electron is a little bit displaced from its partner hole.
This situation is nothing but a finite--momentum BEC, where the
bosons macroscopically occupy a finite $\bqz$ mode, for instance, the
one with $\ell Q = |\ell \bqz| = 1$, Fig.~\ref{fig:bilayer}~(c). Given
such a relation between the momentum $\bqz$ and interlayer coupling,
we also expect that the larger $\ell Q$, the lower the correlation
between the two layers.

These are the key points which motivated us to propose a
finite--momentum BEC of magnetic excitons as a possible ground state
for the bilayer QHS in the intermediate $d/\ell$ region. In the
next two sections, we study the stability of this state at two
different levels of approximation.

As a final remark, we should note that although finite--momentum
BECs have been recently discussed in the context of ultracold Bose
gases (see, for instance, Refs.~\onlinecite{stanescu08,liberto11}), 
our motivation to consider such a phase is mainly due to the
properties of the magnetic exciton as explained above.

\section{One--mode approximation}
\label{sec:onemode}

In this section, we analyze the effective interacting boson model
\eqref{ham-boso} within the so--called Bogoliubov approximation
\cite{fetter} assuming that the ground state is given by a
finite--momentum BEC with the $\bqz = Q\hat{x}$ mode macroscopically
occupied. We hereafter refer to this procedure as the one--mode
approximation.  Although the $\bqz$ mode is not the lowest energy
single--particle boson state, see Fig.~\ref{fig:disp-bosons}, we show
that such a BEC is indeed a stable solution for certain values of $d/\ell$
provided that the excitation spectrum is gapped. Here the boson--boson
interaction potential \eqref{boson-potential} plays an important role
in the stability of this phase. In the following, we
consider $0.1 \le lQ \le 2$ and $0.1\,\ell \le d \le 4\,\ell$.

\begin{figure*}[t]
\centerline{\includegraphics[width=7.4cm]{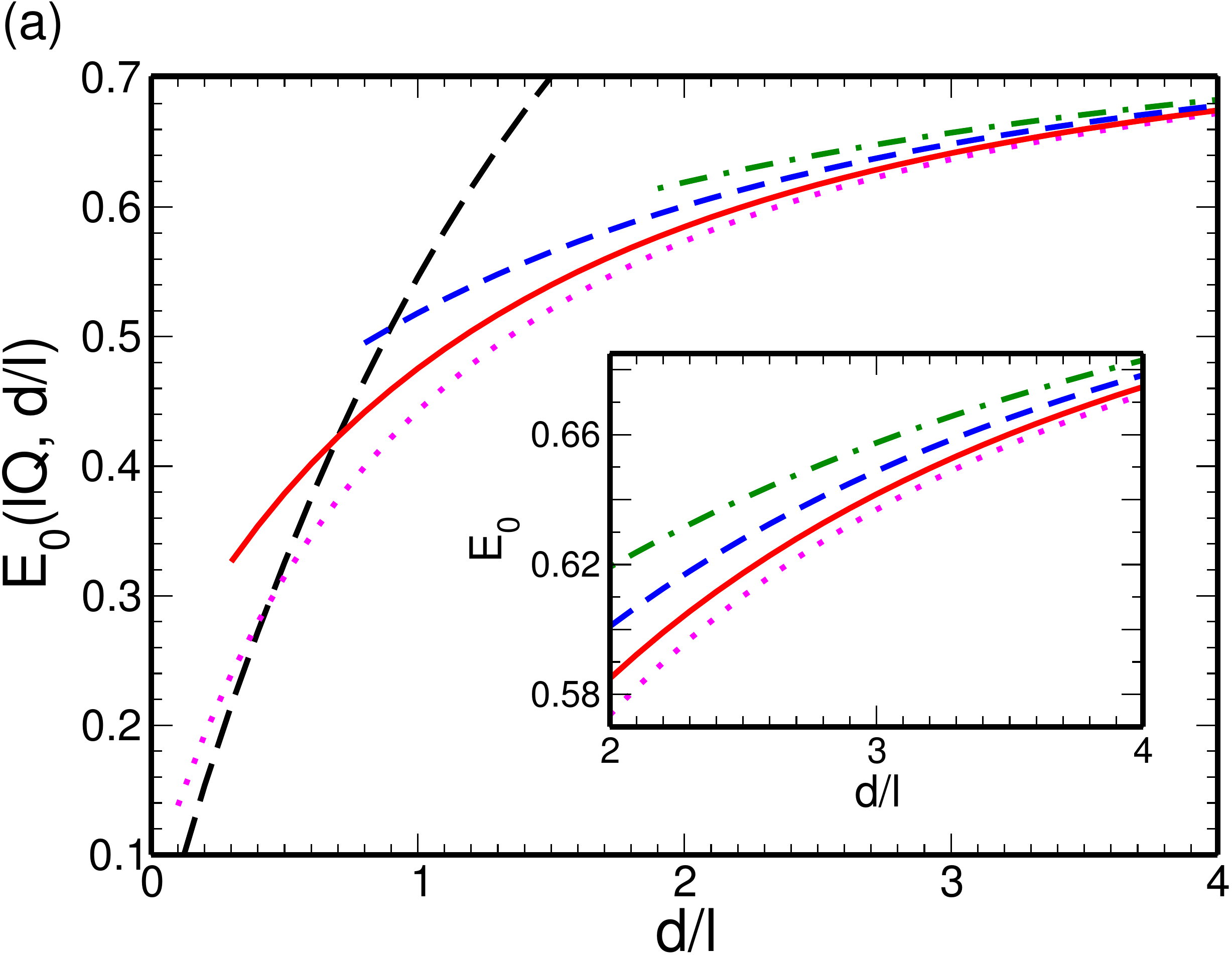}
            \hskip1.5cm
            \includegraphics[width=7.4cm]{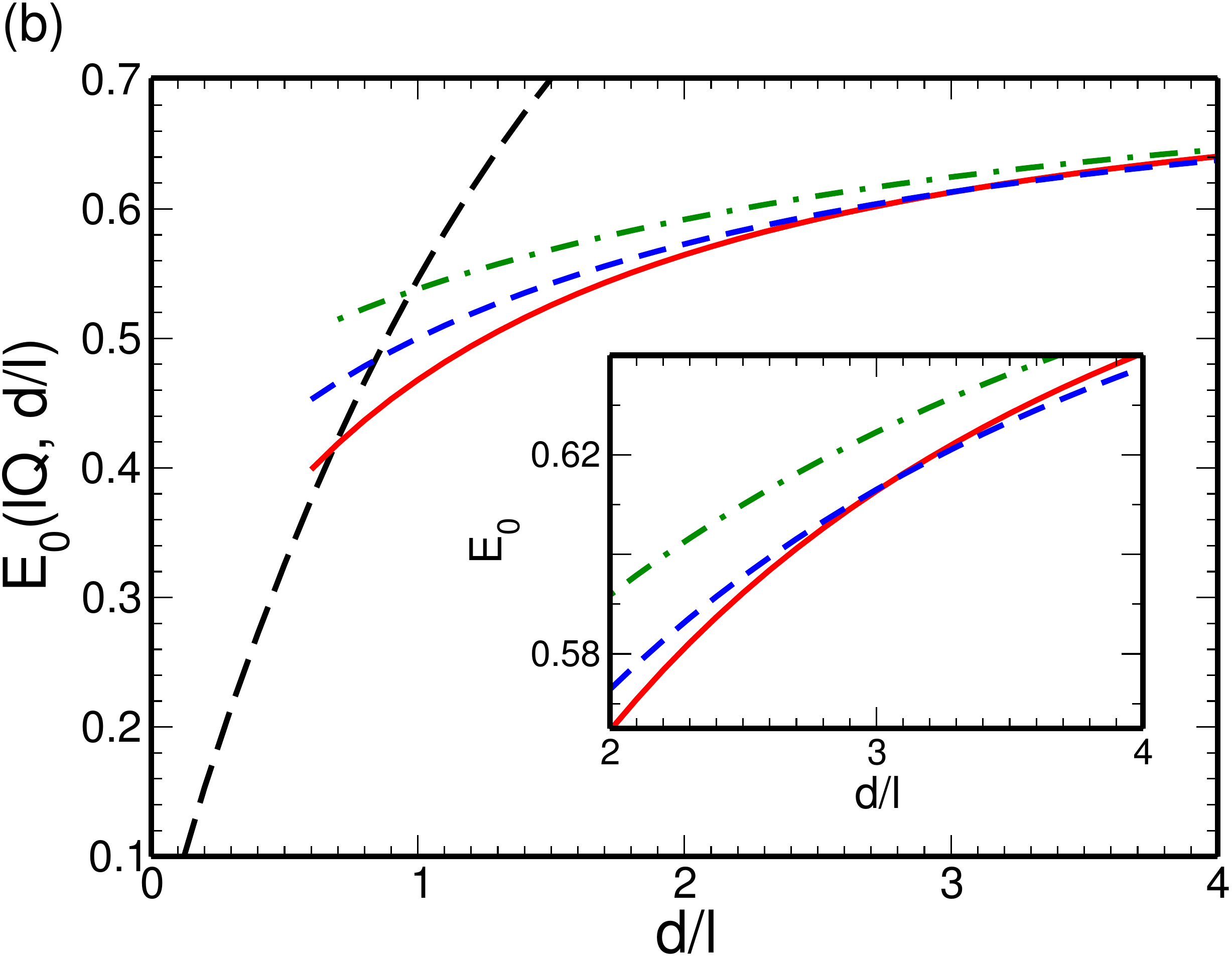}}
\caption{(Color online)  Ground state energy per boson (in units of
  $e^2/\epsilon \ell$) as a function of $d/\ell$: (a) one--mode approximation
  and (b) two--mode approximation. $\ell Q = 0.5$ (dotted magenta
  line), $1$ (solid red line),
  $1.5$ (dashed blue line), and $2.0$ (dot--dashed green line). 
  Long dashed black line: $Q = 0$, one--mode approximation
  with $\mu_0 = 0$, see text for details.
  Inset: details of the corresponding main plots focusing on the
  large $d/\ell$ region.}
\label{fig:egs}
\end{figure*}

Before continuing, some words about the approximation scheme are here
in order: since the single--particle boson energy \eqref{energy-oneboson} 
has cylindrical symmetry, $\omega_\bq = \omega_q$, there is no reason
to expect that the bosons will condense in only one particular momentum $\bqz =
Q\hat{x}$. In principle, the bosons $b$ could even condense in more than
one mode $\bq$ as long as $q=Q$. However, such an approximation is the
simplest one which allows us to verify whether a finite--momentum
BEC of magnetic excitons is indeed a stable phase via quite accurate
and well controlled calculations. This is the idea of the procedure adopted in
this section and in the next one. Later, in Sec.~\ref{sec:discussion}, we
will discuss which of the features found here could be displayed by
the bilayer QHS and also how the results derived from these two
initial considerations could guide us to propose a more elaborated
approximation scheme to study such a finite--momentum BEC.

We start by replacing $b^\dagger_{\bqz} = b_{\bqz} = \langle
b^\dagger_{\bqz} \rangle = \langle b_{\bqz} \rangle \rightarrow
\sqrt{N_0}$ in Eq.~\eqref{ham-k}, where $N_0$ is the (macroscopic)
number of bosons in the $\bqz$ mode. Keeping only terms with two
bosonic operators, one finds after some algebra that
\begin{eqnarray}
  K &=& K_0 + \frac{1}{2}\sum_{\bq\not= 0}
            \left[ \epsilon^+_\bq \, b^\dagger_{\bqz + \bq}b_{\bqz + \bq}
                 + \epsilon^-_\bq \, b_{\bqz - \bq}b^\dagger_{\bqz - \bq}
           \right.
\nonumber \\
        && + \left. \lambda_\bq
         ( b^\dagger_{\bqz + \bq}b^\dagger_{\bqz - \bq}
               +  b_{\bqz - \bq}b_{\bqz + \bq} ) \right],
\label{ham-k2}
\end{eqnarray}
where
\begin{eqnarray}
    K_0 &=& (\omega_{\bqz} - \mu)N_0
            - \frac{1}{2}\sum_{\bq \not= 0}\epsilon^-_\bq,
\nonumber \\
   \epsilon^\pm_\bq &=& \omega_{\bqz \pm \bq} - \mu + \lambda_\bq,
\label{coef-ham-k} \\
&& \nonumber \\
 \lambda_\bq &=& 2N_0v_\bq(\bqz,\bqz).
\nonumber
\end{eqnarray}
The quadratic Hamiltonian \eqref{ham-k2} can be diagonalized with the
aid of the canonical Bogoliubov transformation
\begin{eqnarray}
b^\dagger_{\bqz + \bq} &=& u_\bq a^\dagger_{\bqz + \bq} - v_\bq a_{\bqz - \bq},
\nonumber \\
b^\dagger_{\bqz - \bq} &=& u_\bq a^\dagger_{\bqz - \bq} - v_\bq a_{\bqz + \bq},
\label{bogo-trans}
\end{eqnarray}
which yields
\begin{eqnarray}
  K &=& K_0 + \frac{1}{2}\sum_{\bq\not= 0}
            \left( \Omega^+_\bq \, a^\dagger_{\bqz + \bq}a_{\bqz + \bq}
             + \Omega^-_\bq \, a_{\bqz - \bq}a^\dagger_{\bqz - \bq}
            \right)
\nonumber \\
&=& \bar{K}_0 + \sum_{\bq\not= \bqz}\bar{\Omega}_\bq \, a^\dagger_\bq a_\bq .
\end{eqnarray}
Here
\begin{eqnarray}
   \Omega^\pm_\bq &=& \pm\Delta_\bq  + \Omega_\bq,
\nonumber \\
       \Omega_\bq &=& \sqrt{\epsilon^2_\bq - \lambda^2_\bq},
\nonumber \\
  \epsilon_\bq &=& \frac{1}{2}\left(\epsilon^+_\bq + \epsilon^-_\bq \right),
\;\;\;\;\;
  \Delta_\bq = \frac{1}{2}\left(\epsilon^+_\bq - \epsilon^-_\bq \right),
\nonumber \\
  \bar{K}_0 &=& (\omega_{\bqz} - \mu)N_0  - \frac{1}{2}\sum_{\bq\not= 0}
                \left(\epsilon_\bq - \Omega_\bq \right),
\label{coef-ham-k1}
\end{eqnarray}
the quasiparticle dispersion relation is given by
\begin{equation}
 \bar{\Omega}_\bq \equiv \Omega^+_{\bq - \bqz},
\label{disp-1mode}
\end{equation}
and the Bogoliubov coefficients obey
\begin{eqnarray}
 u^2_\bq &=& \frac{1}{2} + \frac{\epsilon_\bq}{2\Omega_\bq},
\;\;\;\;\;
 v^2_\bq =  -\frac{1}{2} + \frac{\epsilon_\bq}{2\Omega_\bq},
\nonumber \\
 u_\bq v_\bq &=& \frac{\lambda_\bq}{2\Omega_\bq}.
\label{coef-bogo1}
\end{eqnarray}
The chemical potential $\mu$ can be obtained from the saddle point condition
$\partial \bar{K}_0/\partial N_0 = 0$: since
$\partial \epsilon_\bq/\partial N_0 = \partial\lambda_\bq/\partial N_0
= \lambda_\bq /N_0$, one can show that
\begin{equation}
\mu = \omega_{\bqz} + \frac{1}{N_0}\sum_{\bq \not= 0}
          \lambda_\bq v_\bq(v_\bq - u_\bq)
          \equiv \omega_{\bqz} + \mu_0.
\label{eq-mu}
\end{equation}
$N_0$ follows from the conservation (on average) of the total number
of bosons $N_B = \sum_\bq \langle b^\dagger_\bq b_\bq \rangle =
N_\Phi/2 = 1/4\pi \ell^2$: from Eqs.~\eqref{bogo-trans}, one finds that
the relative number of bosons in the condensate is 
\begin{equation}
  n_0 \equiv \frac{N_0}{N_B} = 1 - \sum_{\bq \not= 0}v^2_\bq.
\label{eq-nzero}
\end{equation}
Finally, the ground state energy $E_0(Q,d) = \bar{K}_0 + \mu\langle \hat{N}
\rangle$ reads
\begin{equation}
 \frac{E_0(Q,d)}{N_B} = \frac{\bar{K}_0}{N_B} + \omega_{\bqz} + \mu_0
                      = \omega_{\bqz} + \mu_0(1 - n_0) - I_{01} 
\label{egs1}
\end{equation}
with
$
I_{01} = \frac{1}{2N_B}\sum_{\bq \not= 0}
                   \left(\epsilon_\bq - \Omega_\bq \right).
$
Once $\mu$ and $n_0$ are known for fixed $\bqz$ and $d/\ell$, the
quasiparticle spectrum $\bar{\Omega}_\bq$ and the ground state energy
\eqref{egs1} are completely determined.

\subsection{Zero--momentum BEC}
\label{sec:onemode-zero}

Before proceeding, we would like to briefly recall the results from
our first analysis of the effective boson model \eqref{ham-boso}
reported in Ref.~\onlinecite{doretto06}.

By setting $\bqz = 0$ and $\mu_0 = 0$ in the above equations, we
recover Eqs.~(8) and (9) of Ref.~\onlinecite{doretto06}. The choice
$\mu_0 = 0$, based on the one--loop approximation,\cite{stoof93}
yields a gapless excitation spectrum for the zero--momentum BEC phase
[see Fig.~2(a) from Ref.~\onlinecite{doretto06}], in agreement with the
Goldstone theorem.

We also find that the ground state energy \eqref{egs1} increases with
$d/\ell$, Fig.~\ref{fig:egs} (long dashed black line), and that the
relative number of bosons in the  condensate $n_0$,
Eq.~\eqref{eq-nzero}, decreases rather 
fast as $d/\ell$ increases, Fig.~\ref{fig:nzero}(a). Indeed,
such a behavior of $n_0$ led us to include into the description the
quartic terms in boson operators of the Hamiltonian \eqref{ham-boso}
neglected in the Bogoliubov approximation.\cite{doretto06} Considering these quartic
terms in the so--called Popov approximation,\cite{shi98} we showed that
the self--consistent equations admit solutions only for $d \le
d_{c0} = 0.4\,\ell$. Here, we revisited the problem and perform  
more accurate numerical calculations. We find that $d_{c0} =
0.56\,\ell$, which is even closer to the exact diagonalization 
estimates\cite{simon03,moller08} mentioned
in the Introduction.

\subsection{Finite--momentum BEC}
\label{sec:onemode-finite}

Let us now consider $\ell Q \not= 0$ and discuss numerical solutions of
Eqs.~\eqref{eq-mu}--\eqref{eq-nzero}. It is possible to solve the
self-consistent problem for all values of $\ell Q$ in the considered range
as long as a finite (self-consistently calculated) $\mu_0$ is allowed
and $d$ is larger than a minimum value $d_{\rm min}$. This feature
is exemplified in Fig.~\ref{fig:egs}(a), where we show the
ground state energy \eqref{egs1} as a function of $d/\ell$ for $lQ =
0.5$, $1.0$, $1.5$, and $2.0$. One can see that $d_{\rm min} = 0.1$,
$0.3$, $0.8$, and $1.9\,\ell$ respectively for $d = 0.5$, $1.0$, $1.5$,
and $2.0\,\ell$. Note that the four configurations lie quite close in energy
as $d/\ell$ increases, but the ground state energy curves never cross each
other. This behavior is also observed for all intermediate $\ell Q$
values (not shown here), i.e., $E_0(Q,d)$ increases with $\ell Q$ for
a fixed $d/\ell$. It is clear that a finite--momentum BEC is lower in
energy than the zero--momentum BEC discussed in the previous section
for $d \gtrsim 1.0\,\ell$. Interestingly, the $E_0(Q=0,d)$ and 
$E_0(Q\not=0,d)$ curves cross at a (small) critical layer separation
$d_{c1}$, indicating that a first--order quantum phase
transition from a zero--momentum BEC to a finite--momentum one takes
place at this critical value. Note that for configurations with $0.5
\le \ell Q \le 1.0$, $d_{c1}$ is within the range $0.45\,\ell$ --
$0.7\,\ell$, which includes the (updated) $d_{c0}$ previously
  determined within the Popov approximation in
Ref.~\onlinecite{doretto06}.

One important consequence of a finite $\mu_0$ is that the dispersion
relation of the (neutral) quasiparticles is now gapped. For instance,
in Fig.~\ref{fig:disp-1mode}, we show the excitation spectrum
\eqref{disp-1mode} along some particular momentum directions for the
configuration with $\ell Q=1$ at $d = 1.2\,\ell$. The minimum gap $\Delta$ is
at a momentum $\bq_\Delta = -q_\Delta\hat{x}$, i.e., the angle between
$\bq_\Delta$ and $\bqz$ is equal to $\pi$. For a fixed $lQ$,
$q_\Delta$ continuously increases with $d/\ell$. We also find that,
for a given $lQ$, the gap increases with $d/\ell$ as shown in
Fig.~\ref{fig:gap} (dashed lines). The fact that a gap opens up at
$d_{c1}$ provides further support for a first--order quantum phase
transition at this critical layer separation. Finally, note that  
$\bar{\Omega}_\bq$ has no longer cylindrical symmetry, 
$\bar{\Omega}_\bq \not= \bar{\Omega}_q$, which differs from the
excitation spectrum of the zero--momentum BEC [Fig.~2(a),
Ref.~\onlinecite{doretto06}]. 
This aspect and the peak in $\bar{\Omega}_\bq$ at $\bq
= \bqz$ are artifacts of the oversimplified one--mode approximation.

In order to understand the behavior of the excitation spectrum at
small momentum $\bq$, we should look at the nature of the elementary
excitations. Recall that a boson $b$ has an
internal structure since it corresponds to an electron--hole pair.   
An elementary excitation of the magnetic
exciton BEC can be seen as an electron--hole pair with momentum $\bQ$
which is taken from the condensate, broken and recombined again in a
electron--hole pair but now with a momentum $\bq \not= \bQ$. Apart
from the corrections due to the boson--boson interaction potential,
Eq.~\eqref{boson-potential}, the excitation energy $\bar{\Omega}_\bq$
is related to the difference $\Delta E_b$ between the binding energies
of the pairs with momentum $\bQ$ and $\bq$, namely 
\[ \bar{\Omega}_\bq \sim \Delta E_b =  \omega_\bQ - \omega_\bq, \]  
where $\omega_\bq$ is the dispersion relation of the free bosons,
Eq.~\eqref{energy-oneboson}. Let us firstly consider the
zero--momentum BEC. In this case, the bosons are condensed in the
lowest single--particle energy mode, $\bQ = 0$, and therefore 
$\lim_{\bq\rightarrow 0}(\omega_\bQ - \omega_\bq) = 0$ which yields a
gapless excitation spectrum, i.e., the system displays a Goldstone
mode. On the other hand, in a finite--momentum BEC, the bosons are not
condensed in the lowest single--particle energy mode. This is an
important feature which implies that $\lim_{\bq\rightarrow
0}(\omega_\bQ - \omega_\bq) \not= 0$,  i.e., the Goldstone mode disappears.
In other words, the internal structure of the boson $b$
combined with a macroscopic occupation of a higher energy
single--particle mode leads to the disappearance of 
the Goldstone mode. Such a behavior reminds us of the excitation spectrum of a BCS
superconductor.\cite{tinkham} We will return to this issue in
Sec.~\ref{sec:previous-work}.

\begin{figure}[t]
\centerline{\includegraphics[width=7.5cm]{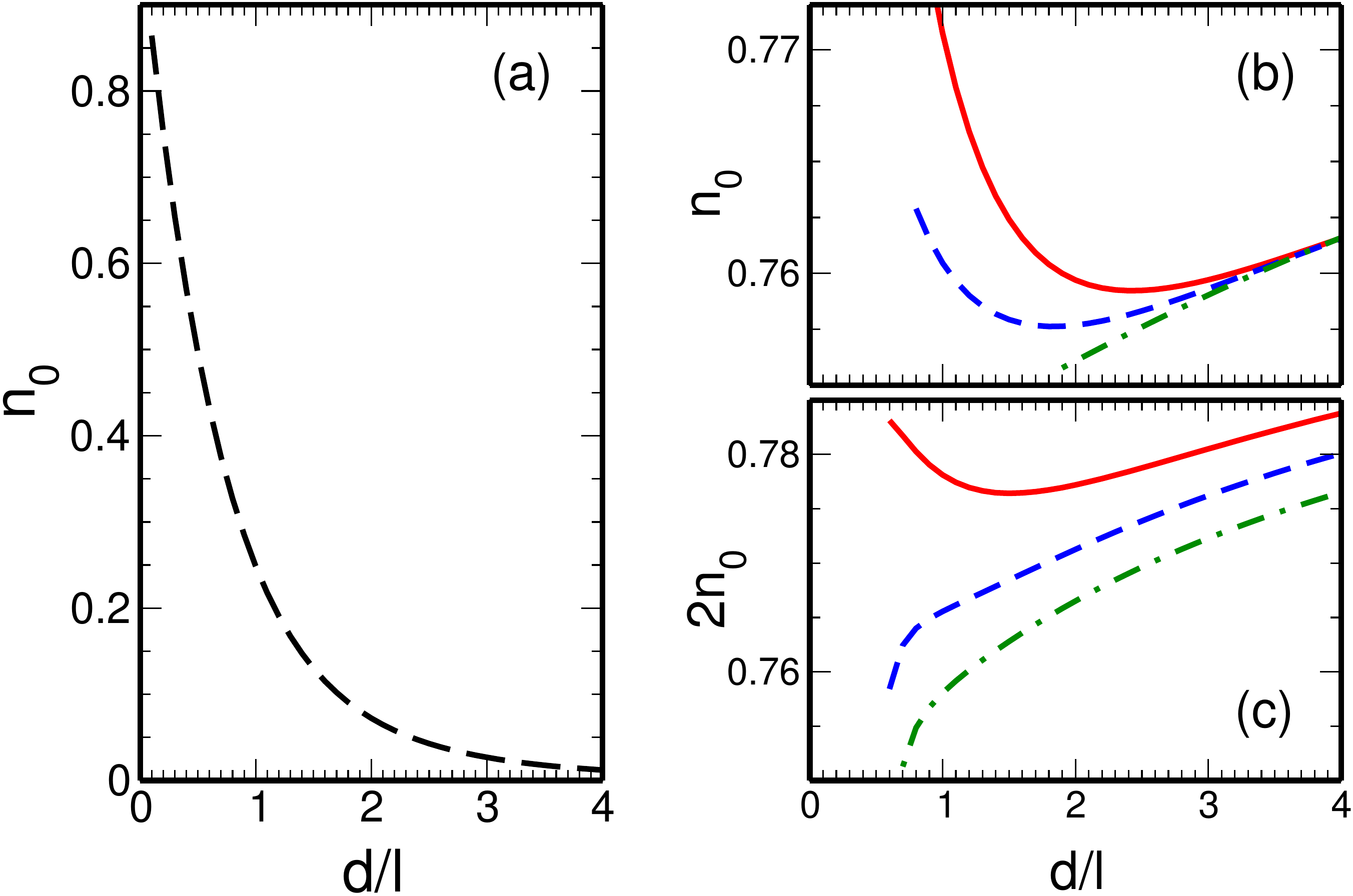}}
\caption{(Color online) Relative number of bosons in the condensate $n_0$,
          Eqs.~\eqref{eq-nzero} and \eqref{eq-nzero2}, as a function of $d/\ell$:
          (a) zero--momentum BEC with $\mu_0 = 0$, see text for details,
          and finite--momentum BEC with $\ell Q = 1.0$ (solid
          red line), $1.5$ (dashed blue line), and $2.0$ (dot--dashed
          green line) within the (b) one--mode
          approximation and (c) two--mode approximation.}
\label{fig:nzero}
\end{figure}

Finally, we find that the relative number of bosons in the condensate
$n_0$, Eq.~\eqref{eq-nzero}, is roughly independent of $d/\ell$ and
close to one. Such an aspect, illustrated in Fig.~\ref{fig:nzero}(b), 
is related to the existence of a finite excitation gap which
reduces quantum fluctuation effects compared with a gapless case (the
zero--momentum BEC). The fact that $n_0 \approx 1$ tell us that the
Bogoliubov approximation is indeed quite reasonable to study a
finite--momentum BEC phase, in contrast with the zero--momentum BEC,
which requires a more involved approximation.

\section{Two--mode approximation}
\label{sec:twomodes}

So far we have considered that the bosons $b$ condense in just one
particular single--particle mode $\bq = Q\hat{x}$. As mentioned in the
previous section, since the single--particle boson dispersion relation
\eqref{energy-oneboson} has cylindrical symmetry, $\omega_\bq =
\omega_q$, the bosons $b$ could, in principle, condense in more than
one mode $\bq$ provided that $q=Q$. In this section, we discuss such a
possibility, in particular, we assume that the
BEC is split into two pieces: both $\bq = \pm\bqz$ modes, with
$\bqz = Q\hat{x}$ and $\ell Q \not= 0$, are now macroscopically occupied.
Again, the Bogoliubov approximation is employed to analyze the effective
boson model \eqref{ham-boso}. We hereafter denote such scheme two--mode
approximation.

\begin{figure}[t]
\centerline{\includegraphics[width=7.5cm]{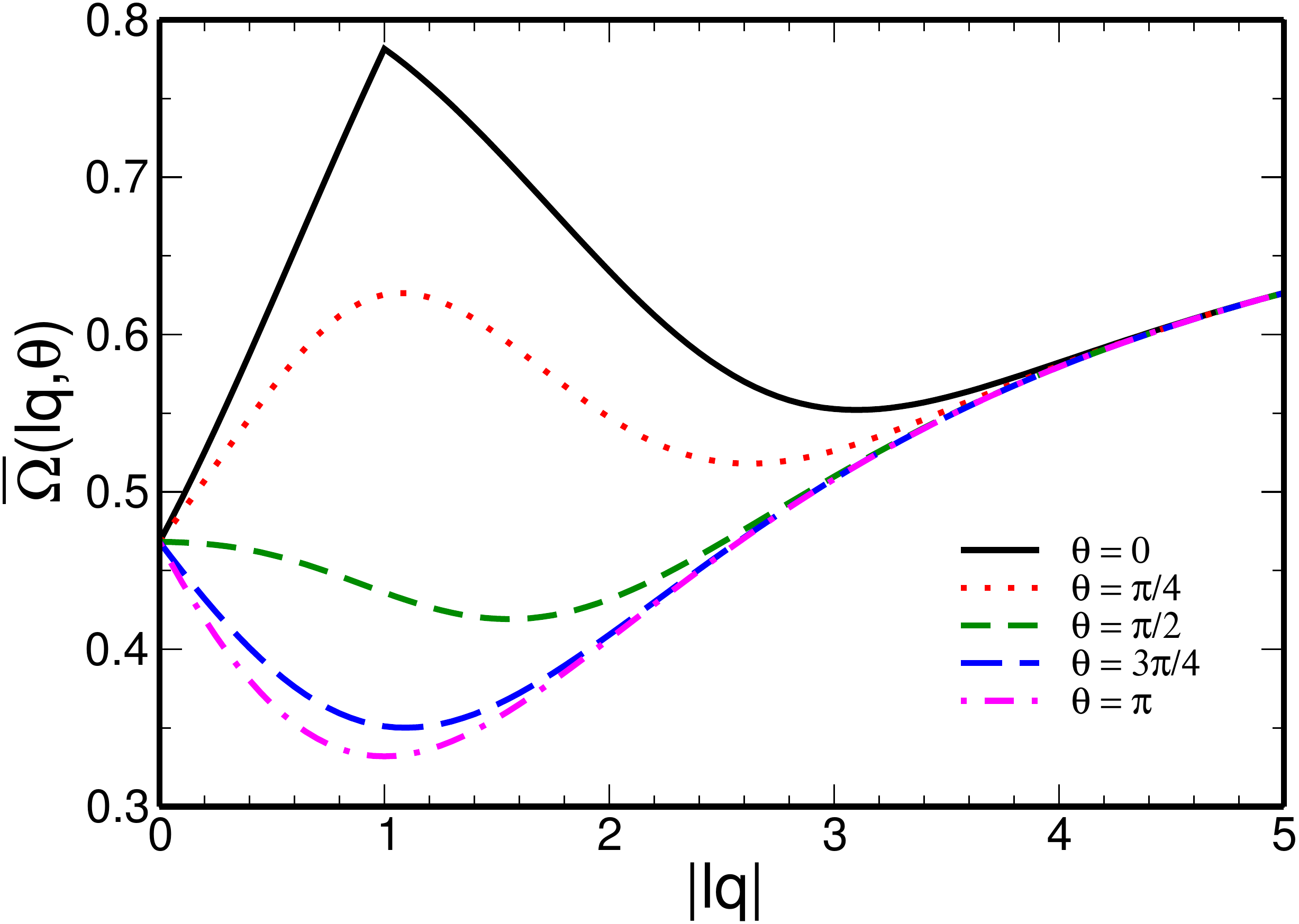}}
\caption{Dispersion relation of the (neutral) quasiparticles,
  Eq.~\eqref{disp-1mode}, (in units of $e^2/\epsilon \ell$) 
  for a finite--momentum BEC with $\ell Q=1$ at $d = 1.2\,\ell$ 
  along some particular momentum directions within the one--mode
  approximation.} 
\label{fig:disp-1mode}
\end{figure}

Here, we basically follow the lines of Sec.~\ref{sec:onemode} and 
start by performing the substitutions
\begin{eqnarray}
  b^\dagger_{\bqz} &=& b_{\bqz} = \langle b^\dagger_{\bqz} \rangle  
            =  \langle b_{\bqz} \rangle \rightarrow \sqrt{N_0}, 
\nonumber \\
  b^\dagger_{-\bqz} &=& b_{-\bqz} = \langle b^\dagger_{-\bqz} \rangle  
            =  \langle b_{-\bqz} \rangle \rightarrow \sqrt{\bar{N}_0}
\label{bogo-subs}
\end{eqnarray}
in Eq.~\eqref{ham-k}. The equivalent of Eq.~\eqref{ham-k2} is now
given by Eq.~\eqref{ham-k4}, see Appendix \ref{ap:details}. In order to diagonalize the Hamiltonian
\eqref{ham-k4}, it is useful to introduce the four component vector
\begin{equation}
  \Psi^\dagger_\bq = \left( b^\dagger_{\bqz + \bq}\;\;
          b^\dagger_{-\bqz + \bq}\; \;
          b_{-\bqz - \bq}\;\;  b_{\bqz - \bq} \right).
\end{equation}
Eq.~\eqref{ham-k4} can then be expressed in matrix form:
\begin{equation}
     K = K_0 + \frac{1}{4}\sum_\bq \Psi^\dagger_\bq \hat{H}_\bq \Psi_\bq,
\label{ham-k4-matrix}
\end{equation}
where the $4\times 4$ matrix $\hat{H}_\bq$ reads
\begin{equation}
 \hat{H}_\bq = \left( \begin{array}{cc}
                          \hat{A}_\bq & \hat{B}_\bq \\
                          \hat{B}_\bq & \hat{A}_\bq
                      \end{array}  \right)
\end{equation}
with the $2\times 2$ matrices $\hat{A}_\bq$ and $\hat{B}_\bq$ given by  
\begin{equation}
 \hat{A}_\bq = \left( \begin{array}{cc}
                           \epsilon_\bq + \Delta_\bq & \gamma_\bq \\
                           \gamma_\bq & \epsilon_\bq - \Delta_\bq
                      \end{array}  \right)
\;\;\;\;{\rm and}\;\;\;\;
 \hat{B}_\bq = \left( \begin{array}{cc}
                             \xi_\bq & \lambda_\bq \\
                             \lambda_\bq & \xi_\bq
                      \end{array}  \right).
\nonumber
\end{equation}
Here, we assume that both condensates have the same number of bosons
and set $\bar{N}_0 = N_0$. The coefficients $\epsilon_\bq$ and
$\Delta_\bq$ are defined in Eq.~\eqref{coef-ham-k1} while
$\gamma_\bq$, $\xi_\bq$, and $\lambda_\bq$ are shown in the Appendix
\ref{ap:details}, see Eqs.~\eqref{coef-ham-k2} and \eqref{coef-ham-k22}.

\begin{figure}[t]
\centerline{\includegraphics[width=7.5cm]{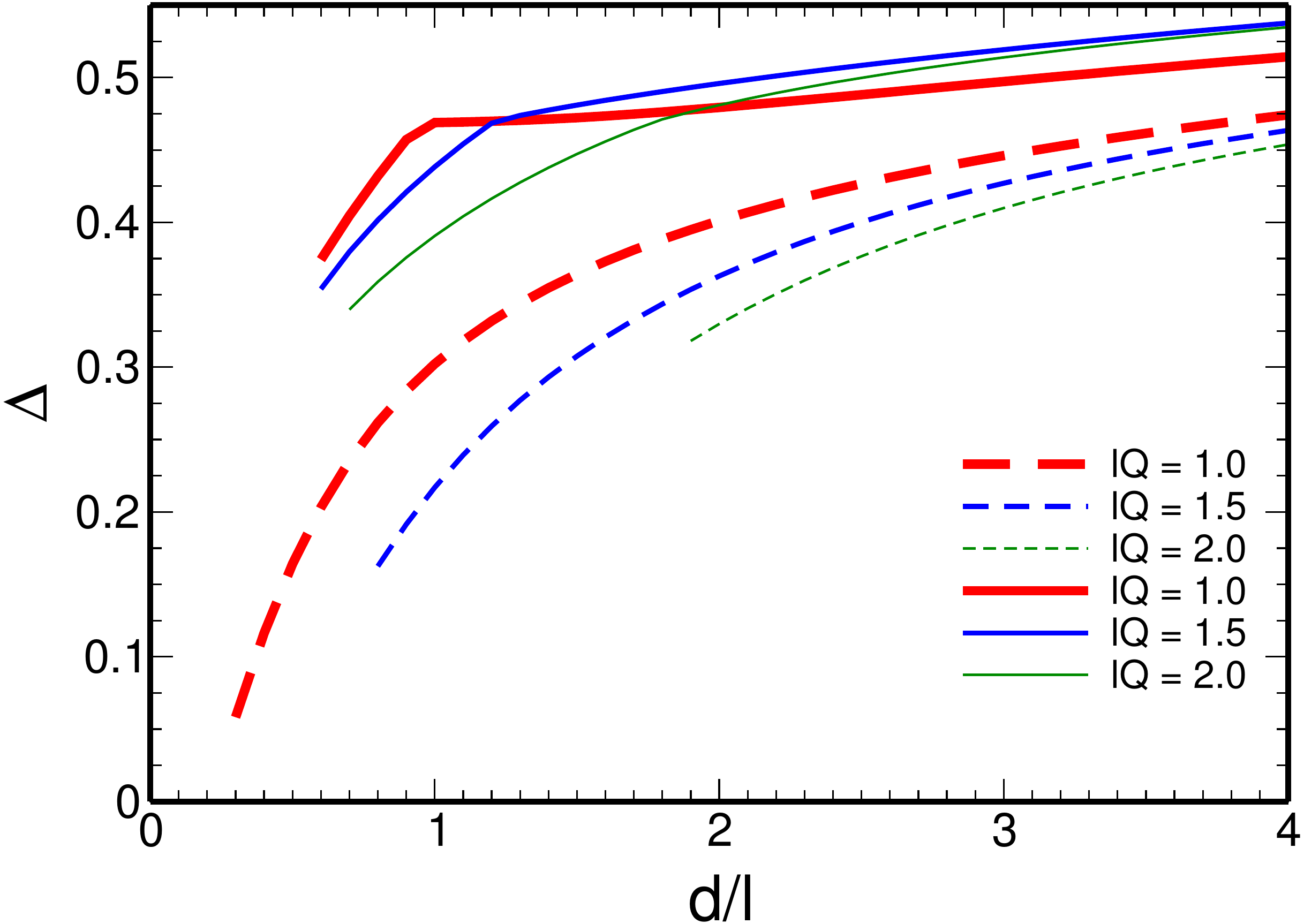}}
\caption{(Color online) Minimum gap energy $\Delta$ (in units of $e^2/\epsilon
  \ell$) of the (neutral) quasiparticle excitations as a function of
  $d/\ell$ for a finite--momentum BEC with $\ell Q = 1$, $1.5$, and $2$.
  Dashed lines: one--mode approximation; solid lines: two--mode
  approximation.}
\label{fig:gap}
\end{figure}

The diagonalization of the $4\times 4$ problem \eqref{ham-k4-matrix}
is more involved than the $2\times 2$ one corresponding to
Eq.~\eqref{ham-k2}. Therefore, it is more convenient here to use
the procedure described in Ref.~\onlinecite{blaizot}: Since we are
dealing with a bosonic system, instead of $\hat{H}_\bq$, one should
diagonalize
\begin{equation}
 \hat{I}_B\hat{H}_\bq ,
 \;\;\;\;\;\;\; {\rm with}
 \;\;\;\;\;\;\;
 \hat{I}_B = \left( \begin{array}{cc}
                      \hat{I} & 0 \\ 0 & -\hat{I}
                      \end{array}  \right).
\label{aux-ham}
\end{equation}
The (positive) eigenvalues of the matrix \eqref{aux-ham} are
\begin{equation}
 \Omega^\pm_\bq = \sqrt{C_\bq \pm 2D_\bq},
\label{disp-2modes}
\end{equation}
where
\begin{eqnarray}
    C_q &=& \epsilon^2_\bq + \Delta^2_\bq + \gamma^2_\bq
          - \lambda^2_\bq  - \xi^2_\bq,
\nonumber \\
    D_q &=& \left[  \Delta^2_\bq\left(\epsilon^2_\bq - \lambda^2_\bq \right)
             + \left(\gamma_\bq\epsilon_\bq - \lambda_\bq\xi_\bq
                       \right)^2\right]^{1/2}.
\end{eqnarray}
Eq.~\eqref{ham-k4-matrix} then acquires the form
\begin{equation}
     K = K_0 + \frac{1}{4}\sum_\bq \Phi^\dagger_\bq \hat{H}'_\bq \Phi_\bq,
\label{ham-k4-matrix2}
\end{equation}
where the $4\times 4$ matrix $\hat{H}'_\bq$ reads
\begin{equation}
 \hat{H}'_\bq = \left( \begin{array}{cc}
                          \hat{h}_\bq & 0 \\
                            0         & \hat{h}_\bq
                         \end{array}  \right)
\;\;\;\;\; {\rm with}\;\;\;\;\;
 \hat{h}_\bq = \left( \begin{array}{cc}
                           \Omega^+_\bq &  0 \\
                                    0   & \Omega^-_\bq
                      \end{array}  \right)
\end{equation}
and the new four component vector $\Phi^\dagger_\bq$ is given by
\begin{equation}
      \Phi^\dagger_\bq = \left( a^\dagger_{\bqz + \bq} \;\;
                               a^\dagger_{-\bqz + \bq} \;\;
                               a_{-\bqz - \bq} \;\;
                               a_{\bqz - \bq} \right).
\label{def-a-bosons}
\end{equation}
The relation between the two set of bosonic operators
$a_{\pm\bqz\pm\bq}$ and $b_{\pm\bqz\pm\bq}$ is
\begin{equation}
 \Phi_\bq = \hat{M}_\bq\Psi_\bq,
 \;\;\;\;\;\;\; {\rm where} \;\;\;\;\;\;\;
 \hat{M}_\bq = \left( \begin{array}{cc}
                            \hat{U}_\bq & \hat{V}_\bq \\
                            \hat{V}_\bq & \hat{U}_\bq
                            \end{array}  \right)
\label{m-matrix}
\end{equation}
with $\hat{U}_\bq$ and $\hat{V}_\bq$ being $2\times 2$ matrices,
\begin{equation}
 \hat{U}_\bq = \left( \begin{array}{cc}
                             u_1(\bq) & u_3(\bq) \\
                             u_2(\bq) & u_4(\bq)
                             \end{array}  \right),
\;\;\;\;\;\;\;
 \hat{V}_\bq = \left( \begin{array}{cc}
                             v_1(\bq) & v_3(\bq) \\
                             v_2(\bq) & v_4(\bq)
                             \end{array}  \right),
\nonumber
\end{equation}
whose elements are the Bogoliubov coefficients.
The complete expressions of the Bogoliubov coefficients $u_i(\bq)$ and
$v_i(\bq)$ are quite long and they can be found in the
Appendix~\ref{ap:details}.

Eq.~\eqref{ham-k4-matrix2} can be rewritten as
\begin{equation}
  K = \bar{K}_0 + \sum_{\bq\not= \pm\bqz} \bar{\Omega}_\bq \,
      a^\dagger_\bq a_\bq ,
\end{equation}
where the quasiparticle energy $\bar{\Omega}_\bq$ reads
\begin{equation}
  \bar{\Omega}_\bq =  \frac{1}{4}\left(\Omega^+_{\bqz + \bq}
              + \Omega^+_{-\bqz + \bq}+\Omega^-_{\bqz + \bq}
              + \Omega^-_{-\bqz + \bq}   \right)
\label{disp-2modes2}
\end{equation}
and
\begin{eqnarray}
 \bar{K}_0 &=& 2N^2_0v_{2\bqz}(\bqz,\bqz) + 2N_0(\omega_{\bqz} - \mu)
\nonumber  \\
&& \nonumber \\
     &+& \frac{1}{4}\sum_\bq
         \left(\Omega^+_\bq + \Omega^-_\bq - \epsilon^+_\bq
               - \epsilon^-_\bq \right).
\end{eqnarray}
Again, from the saddle point condition $\partial \bar{K}_0/\partial
N_0 = 0$, the chemical potential $\mu$ can be calculated: since from
Eqs~\eqref{coef-ham-k2} with $\bar{N}_0 = N_0$ we have
\[
\partial \epsilon^\pm_\bq/\partial N_0 = 
\partial \lambda_\bq/\partial N_0 = \lambda_\bq/N_0,
\;\;\;\;\;\;\;
\partial \Delta_\bq/\partial N_0 = 0,
\]
\[
\partial \gamma_\bq/\partial N_0 = \gamma_\bq/N_0,
\;\;\;\;\;\;\;
\partial \xi_\bq/\partial N_0 = \xi_\bq/N_0,
\]
after some algebra, we find that
\begin{equation}
   \mu = \omega_{\bqz} + \mu_0 + 2N_0 v_{2\bqz}(\bqz,\bqz)
         \equiv \omega_{\bqz} + \mu_0 + \mu_1.
\label{eq-mu2}
\end{equation}
The quantity $\mu_0$, see Eq.~\eqref{mu0-2modes}, 
is different from the one--mode approximation expression,
Eq.~\eqref{eq-mu}. From the conservation (on average) of the total
number of bosons $N_B = \sum_\bq \langle b^\dagger_\bq b_\bq \rangle =
2N_0 + \sum_{\bq \not= \pm \bqz} \langle b^\dagger_\bq b_\bq \rangle$,
it follows that the relative number of bosons in the condensate $n_0$
is given by 
\begin{equation}
 n_0 \equiv \frac{N_0}{N_B} = \frac{1}{2}
            - \frac{1}{4}\sum_{i=1}^4\sum_\bq v^2_i(\bq).
\label{eq-nzero2}
\end{equation}
Finally, it is easy to see that the ground state energy
$E_0(Q,d) = \bar{K}_0 + \mu\langle \hat{N} \rangle$ reads
\begin{eqnarray}
\frac{E_0(Q,d)}{N_B} &=& \frac{\bar{K}_0}{N_B} + \omega_{\bqz}
                    + \mu_0 + \mu_1
\nonumber \\
&& \nonumber \\
        &=& \omega_{\bqz} + \mu_0(1 - 2n_0) + \mu_1(1 - n_0) 
            - I_{02}\;\;\;\;\;
\label{egs2}
\end{eqnarray}
with
$
I_{02} = \frac{1}{4N_B}\sum_\bq \left(\epsilon^+_\bq + \epsilon^-_\bq 
           - \Omega^+_\bq - \Omega^-_\bq  \right).
$
Similarly to the one--mode approximation, Sec.~\ref{sec:onemode}, we
numerically solve the self--consistent Eqs.~\eqref{eq-mu2} and
\eqref{eq-nzero2} and determine $n_0$ and $\mu$ for fixed $\ell Q$ and
$d/\ell$.

\begin{figure}[t]
\centerline{\includegraphics[width=7.5cm]{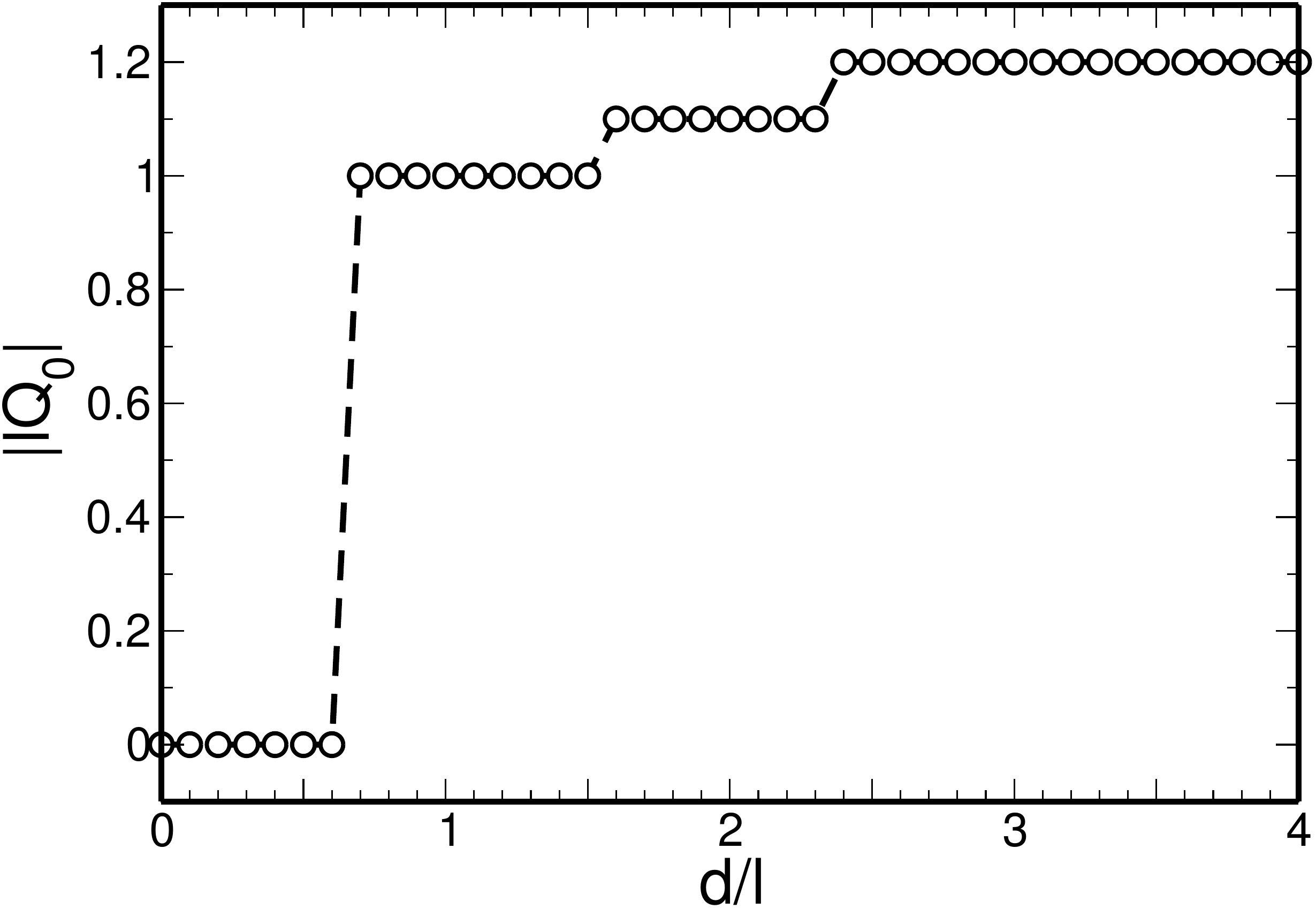}}
\caption{Magnitude of the momentum $\ell\bqz_0$ corresponding to the
  lowest energy configuration for a fixed $d/\ell$, two--mode approximation.
  Dashed line is a guide to the eyes.}
\label{fig:qzero}
\end{figure}

In Fig.~\ref{fig:egs}(b), we plot the ground state energy
\eqref{egs2} as a function of $d/\ell$ for three different
configurations, the ones with $\ell Q = 1.0$, $1.5$, and $2.0$. Likewise
the one--mode approximation, the self--consistent equations can be
solved only for $d$ larger than a minimum value $d_{\rm min}$. 
However, we now have $d_{\rm min} = 0.6,$ $0.6$, and $0.7\,\ell$ respectively
for $\ell Q = 1.0$, $1.5$, and $2.0$, which differ from the one--mode
approximation results. Comparing the ground state energies obtained
with both one--mode and two--mode approximations for a given $\ell
Q$, we clearly see that the latter is lower than the former: since
$\omega_\bq = \omega_q$, macroscopic occupation of both $\pm\bqz$
modes are equally likely. The system then profits from this fact by
splitting the condensate into $n=2$ equal pieces, binding them and
lowering the total energy. Again, a finite--momentum BEC is more
favorable than a zero--momentum one for $d \gtrsim 1.0\,\ell$ and the $E_0(Q
\not= 0,d)$ and the $E_0(Q=0,d)$ [one--mode approximation with
$\mu_0=0$] curves cross at small $d$. In particular, $E_0(lQ=1,d)$ and
$E_0(Q=0,d)$ cross at $d_{c1} \approx 0.68\,\ell$, in good agreement
with the result derived in the previous section. Therefore, both
one--mode and two--mode approximations indicate that a first--order
quantum phase transition may occur at small $d/\ell$.

Concerning the large $d$ region, we again find that the
configurations with $\ell Q \not= 0$ are quite close in energy but now,
differently from the one--mode approximation, the different $E_0(Q
\not= 0,d)$ curves cross each other. For instance,
$E_0(\ell Q=1,d)$ and $E_0(\ell Q=1.5,d)$ cross at $d \approx
3.1\,\ell$, see inset Fig.~\ref{fig:egs}(b). Indeed, we find
several crossings between the different ground state energy curves for
$1.0 \le \ell Q \le 2.0$ and $0.7\,\ell \le d \le 4\,\ell$. In
particular, the magnitude of the
momentum $|\ell\bqz_0|$ corresponding to the lowest energy configuration
for a given $d/\ell$ is shown in Fig.~\ref{fig:qzero}. The fact that
$\ell Q_0$ increases with $d/\ell$ corroborates the scenario proposed in
Sec.~\ref{sec:ansatz} that the larger $\ell Q$, the lower the correlation
between the two layers. Note that the one--mode approximation is not
enough to capture such a behavior. Moreover, the results also indicate
that another first--order quantum phase transition may take place at
larger $d$, from one finite--momentum BEC with small $\ell Q_0$ to
another one with a larger $\ell Q_0$. In particular, the transition
$\ell Q_0 = 1.0$ $\rightarrow$ $\ell Q_0 = 1.1$ occurs at $d_{c2} = 1.6\,\ell$, which
is very close to the critical layer separation where the
incompressible--compressible phase transition is experimentally
observed.\cite{exp-bilayer} Finally, we should mention that solutions
for the $\ell Q=0.8$ and $0.9$ configurations are also possible but,
since they are very close in energy to the $\ell Q = 1.0$
configuration, we decided to neglected them in the above discussion.

\begin{figure}[!t]
\vskip-0.7cm
\centerline{\includegraphics[width=7.2cm]{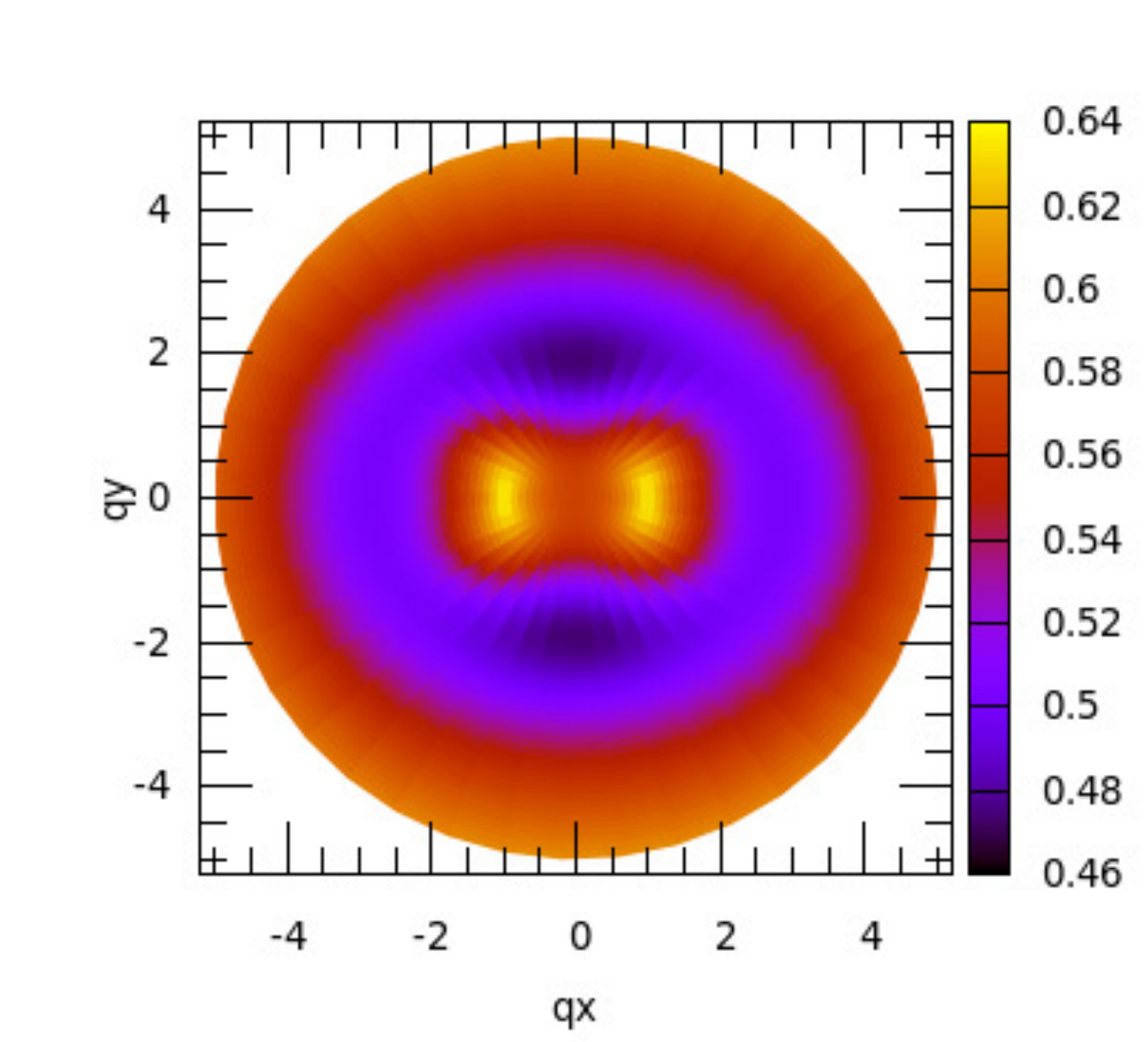}}
\caption{(Color online) Contour plot of the quasiparticle dispersion
  relation $\bar{\Omega}_\bq$, Eq.~\eqref{disp-2modes2}, (in units of
  $e^2/\epsilon\ell$) for a finite--momentum BEC with $\ell Q = 1$ at $d
  = 1.5\,\ell$, two--mode approximation.} 
\label{fig:contour}
\end{figure}

\begin{figure*}[t]
\centerline{\includegraphics[width=5.7cm]{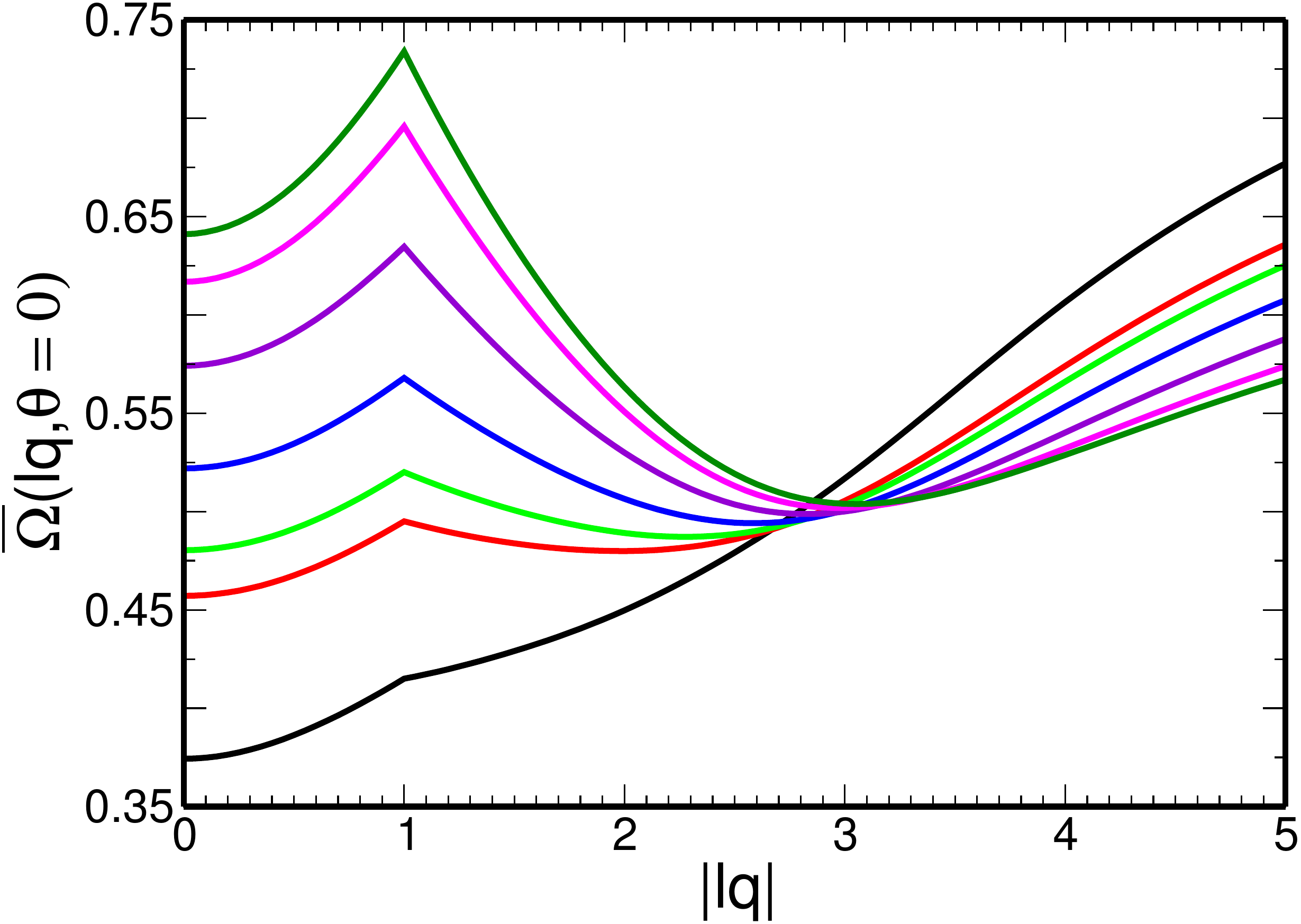}
            \hskip0.2cm
            \includegraphics[width=5.7cm]{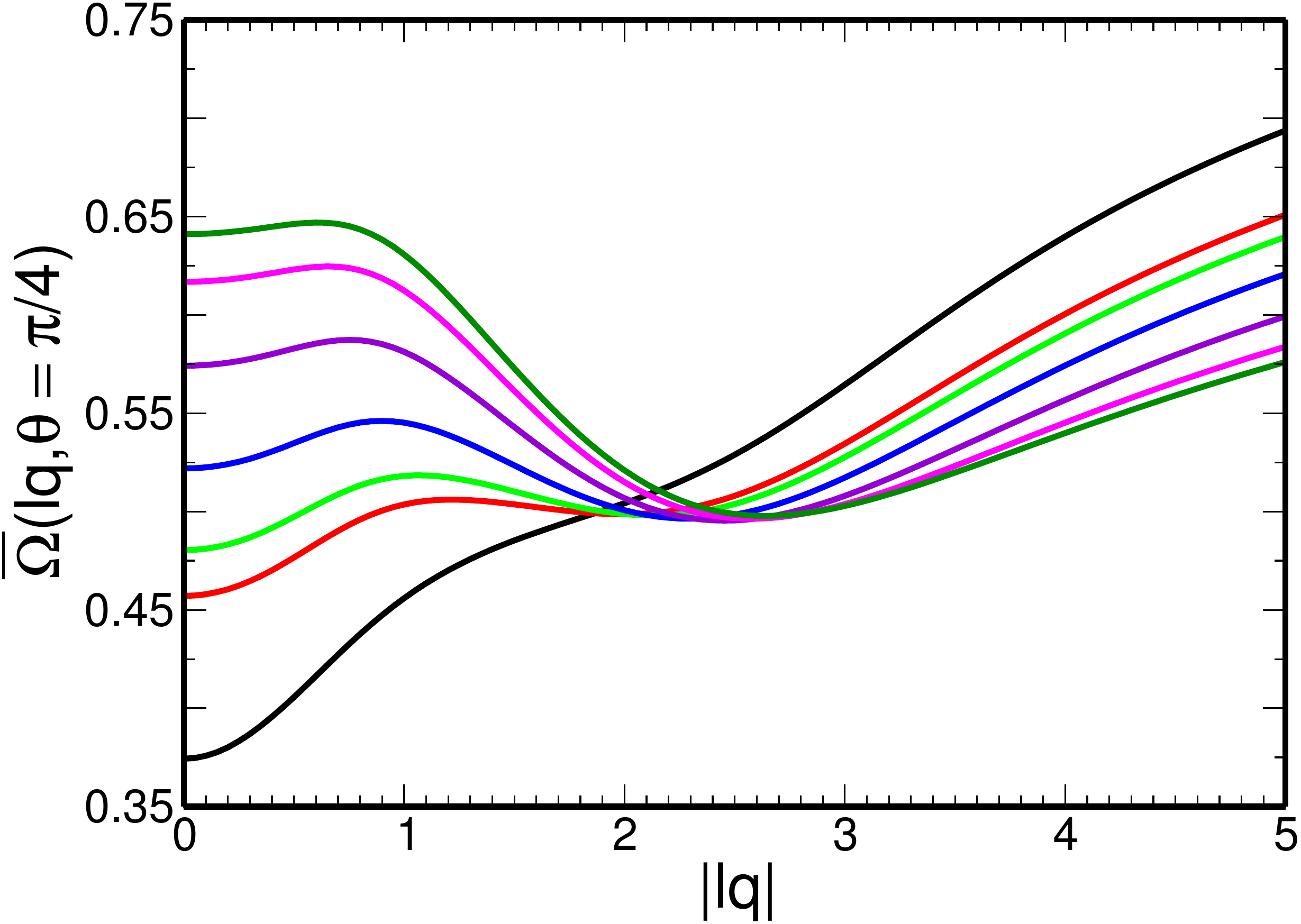}
            \hskip0.2cm
            \includegraphics[width=5.7cm]{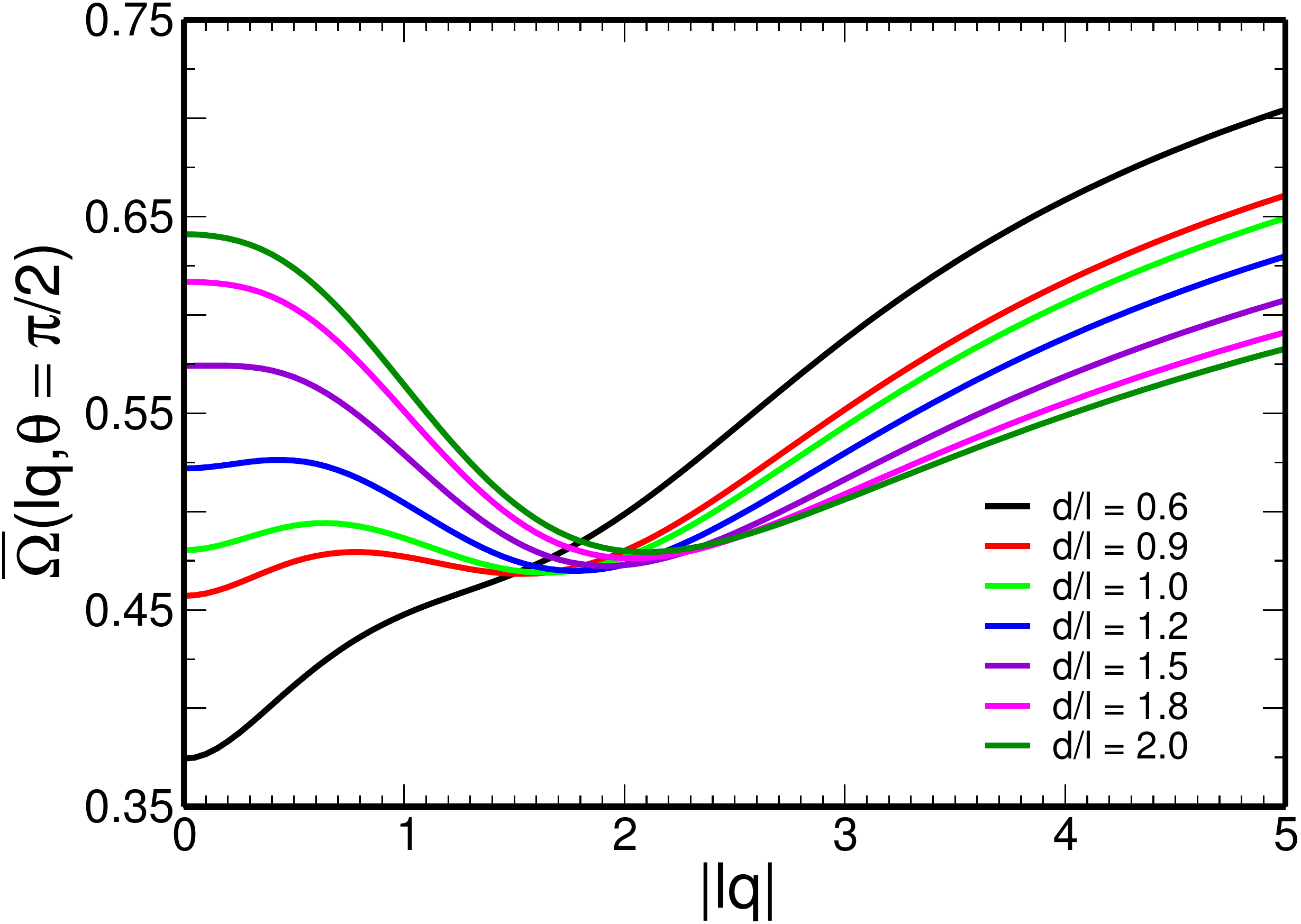}}
\vskip0.7cm
\centerline{\includegraphics[width=5.7cm]{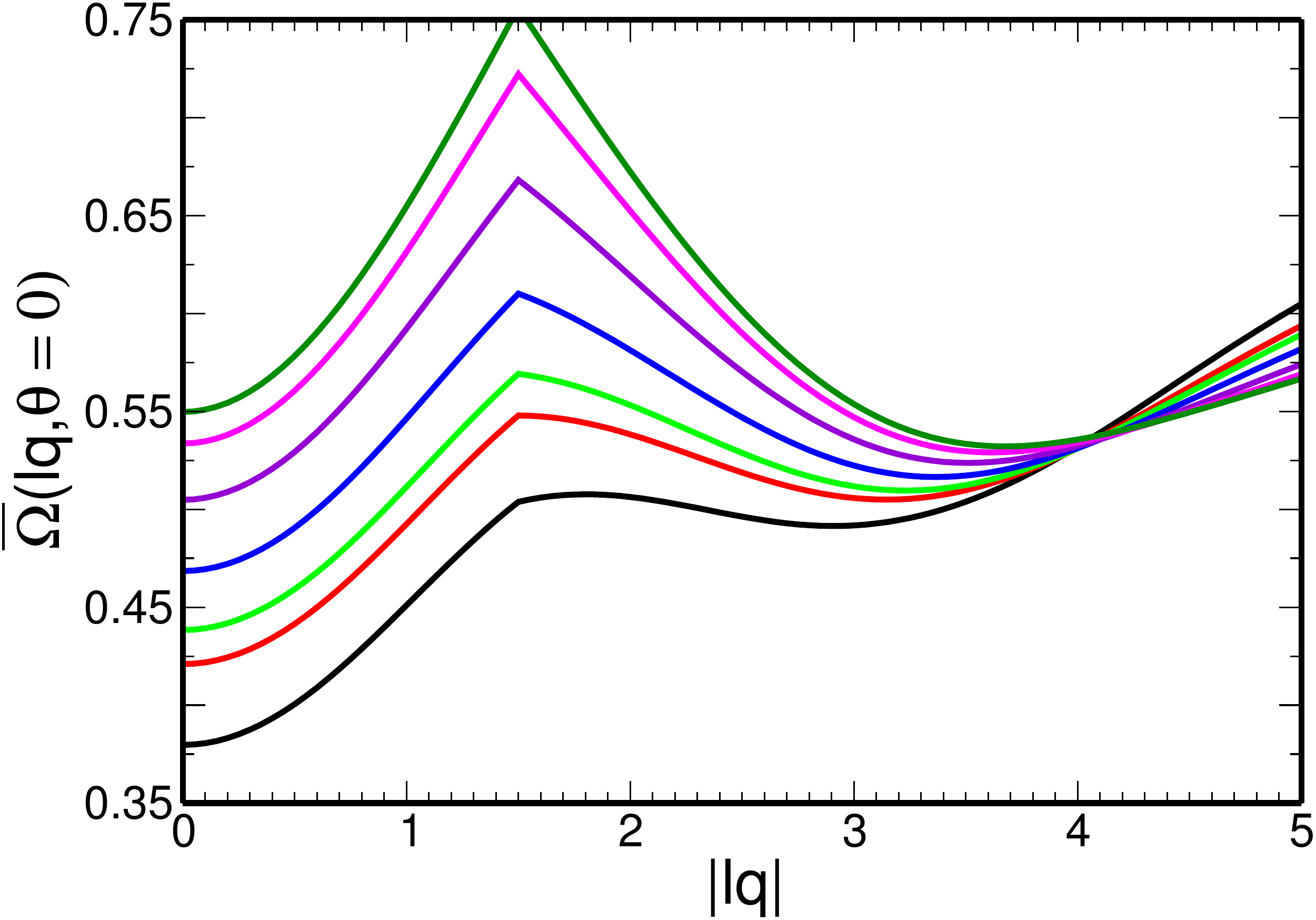}
            \hskip0.2cm
            \includegraphics[width=5.7cm]{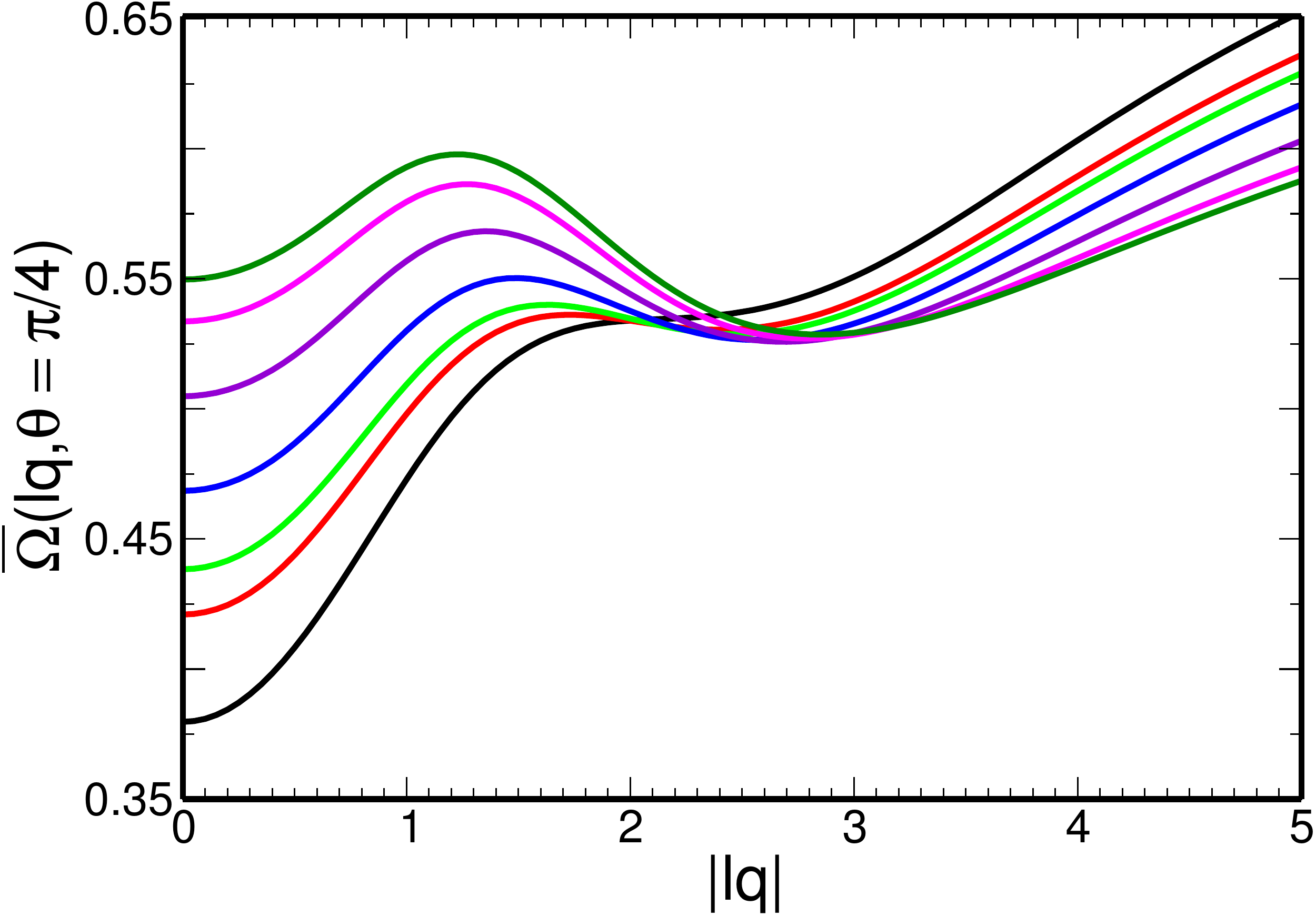}
            \hskip0.2cm
            \includegraphics[width=5.7cm]{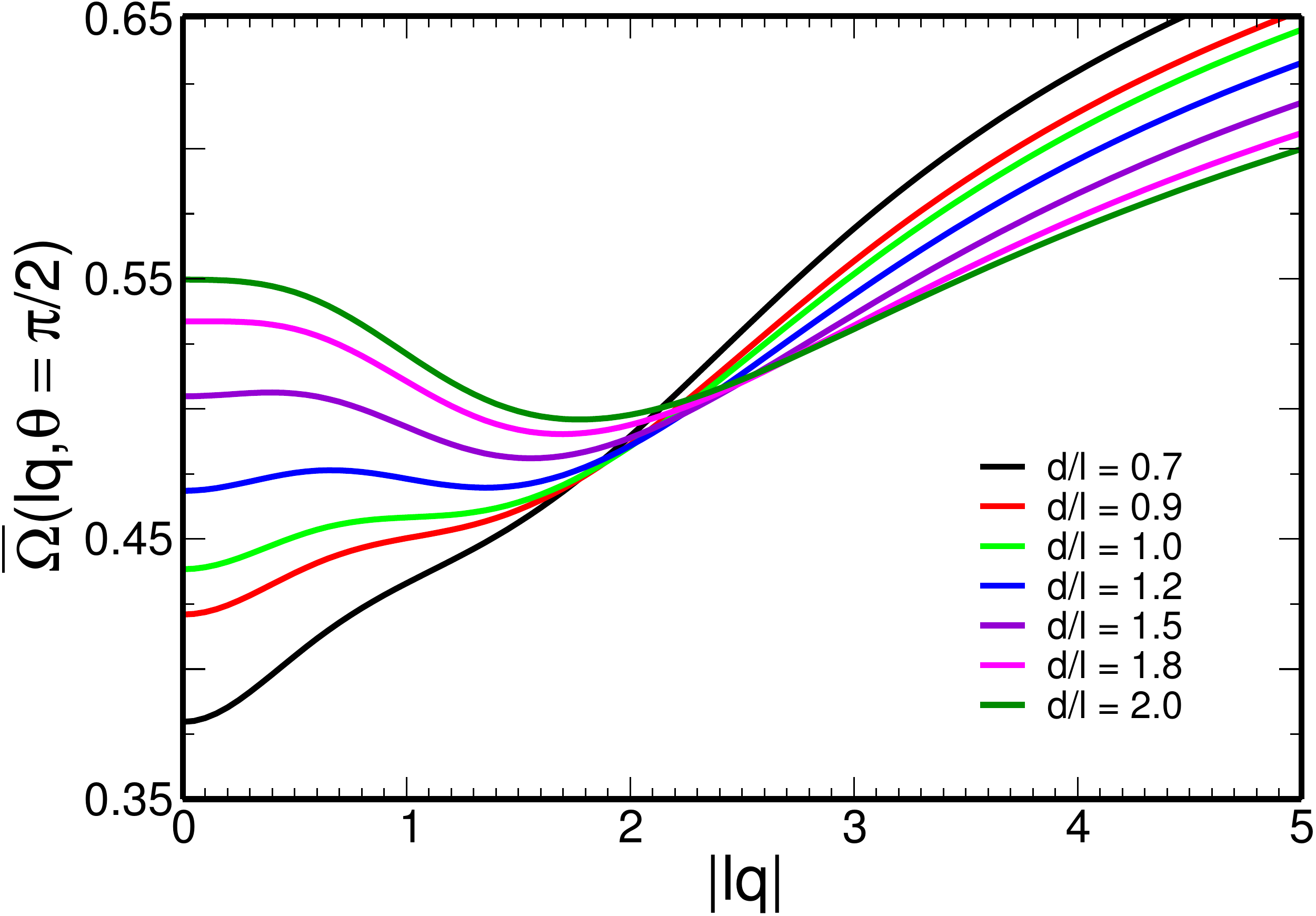}}
\caption{(Color online) Quasiparticle dispersion relation $\bar{\Omega}_\bq$,
  Eq.~\eqref{disp-2modes2}, (in units of $e^2/\epsilon\ell$) along some
  particular momentum directions for several values of the ratio
  $d/\ell$. Finite--momentum BEC with $\ell Q = 1.0$ (upper row) and
  $\ell Q = 1.5$ (lower row), two--mode approximation ($d/\ell$
  increases from bottom to top at $\ell q = 0$).}
\label{fig:disp-2modes}
\end{figure*}

Differently from the one--mode approximation, here the inversion
symmetry of the excitation spectrum $\bar{\Omega}_\bq$,
Eq.~\eqref{disp-2modes2}, is preserved as exemplified in
Fig.~\ref{fig:contour} for the finite--momentum BEC with $\ell Q = 1.0$ at
$d = 1.5\ell$. Again, a finite (self-consistently determined) $\mu_0$
leads to a gapped excitation spectrum. Note that the minimum gap, which is
larger than the corresponding one determined within the one--mode
approximation, increases with $d/\ell$ for a fixed $\ell Q$,
Fig.~\ref{fig:gap} (solid lines). Interestingly, for small $d/\ell$, the
minimum gap is located at the origin ($\bq_\Delta = 0$) but, as $d/\ell$
increases, the position of the minimum gap abruptly changes to $\bq_\Delta =
q_\Delta\hat{y}$ (the angle between $\bq_\Delta$ and $\bqz$ is now
$\pi/2$, in contrast with the one--mode approximation result) and then
$\ell q_\Delta$ continuously increases with $d/\ell$. Such a behavior
is exemplified in Figs.~\ref{fig:disp-2modes} for the $\ell Q = 1.0$
(upper row) and $1.5$ (lower row) configurations. Also, the kinks
observed in Fig.~\ref{fig:gap} (solid lines) are signatures of this
abrupt change in $\bq_\Delta$. Finally, some words about the
singularities of the excitation spectrum are here in order: we believe
that the peaks in $\bar{\Omega}_\bq$ at $\pm \bqz$ (also found in the
one--mode approximation) might be an
artifact of the two--mode approximation and that they may disappear
as we increases the number of components $n$ (even) of the
finite--momentum condensate. We will return to this point in
Sec.~\ref{sec:consequences}. 

Concerning the relative number of bosons in the two condensate pieces
$2n_0$, Fig.~\ref{fig:nzero}(c), we can see that its
behavior is similar to the one found in the previous section: it is
almost independent of $d/\ell$ and $2n_0$ is close to one. The latter
indicates that the Bogoliubov approximation is indeed appropriate to
study the finite--momentum BEC even if the condensate is split into
more than one piece.

As a final remark, we would like to comment on the fact that the
results found in this section seem to contradict Nozi\`eres,\cite{nozieres} 
who argued that a fragmentation of the condensate into two pieces 
costs a macroscopic extensive exchange energy and therefore it is not
favorable. A careful analysis shows that there is no contradiction. 
Let us denote $E_{01}$ and $E_{02}$ the ground state energy respectively
obtained within the one--mode [Eq.~\eqref{egs1}] and two--mode
[Eq.~\eqref{egs2}] approximations. Using the fact that $n_0$ [one--mode
 approximation, Eq.~\eqref{eq-nzero}] is roughly equal to $2n_0$
[two--mode approximation, Eq.~\eqref{eq-nzero2}], we have
\[
  E_{02} - E_{01} \approx \mu_1(1 - n_0) - I_{02} + I_{01}. 
\]
Comparing the above equation with Eq.~(4) from Ref.~\onlinecite{nozieres}, 
we realize that Nozi\`eres considerations only take into account the
first term in the above equation and completely neglect the other
terms which, as we have seen, provide important corrections. 
In particular, for the bilayer QHS, symmetry considerations also indicate
that a BEC split into two parts has lower energy than a single
condensate: recall that the excitation spectrum obtained within the
two--mode approximation is more symmetrical than the one derived in
the one--mode approximation.

\section{Properties of a BEC of magnetic excitons}
\label{sec:properties}

It is easy to see from Eq.~\eqref{rho-boso} that
\begin{equation}
\langle \rho(\bk) \rangle = \delta_{k,0}N_\Phi,
\label{exp-rho}
\end{equation}
which is valid for both one and two--mode approximations
regardless the value of $lQ$. 
Eq.~\eqref{exp-rho} implies that, in principle, the BEC of magnetic
excitons is an homogeneous phase (see discussion at the end of the section).
Concerning the expectation value of the 
$\hat{z}$--component of the pseudospin density operator
\eqref{sz-boso}, one shows (two--mode approximation) 
\begin{equation}
 \langle S_z(\bk) \rangle = \sqrt{N_0\bar{N}_0}\exp(-|\ell Q|^2/4)
                           \left(\delta_{\bk,-2\bqz} + \delta_{\bk,2\bqz} \right).
\label{exp-sz}
\end{equation}
Note that \eqref{exp-sz} vanishes within the one--mode approximation
since $\bar{N}_0 \rightarrow 0$.

Further insight into a BEC of magnetic excitons can be obtained by 
looking at the pair correlation function, which is defined
as\cite{mahan} 
\begin{equation}
  g(\br) - 1 = \frac{1}{N}\sum_\bq e^{-i\bq\cdot\br}[S(\bq) - 1],
\label{eq-pairfunction}
\end{equation}
where the static structure factor is given by
\begin{equation}
  S(\bq) = \frac{1}{N}\langle \rho(-\bq)\rho(\bq)\rangle - 
              N\delta_{q,0}
\label{eq-pair-s}
\end{equation}
with $\rho(\bq)$ being the Fourier transform of the electron density
operator. The pair correlation function basically tell us the
probability of finding an electron at the position $\br$ giving that
there is another one at the origin. The analytical expression of
$g(\br)$, both at the one and two--mode approximations, is quite
lengthy and can be found in the Appendix~\ref{ap:pair}. Here, we just
comment on its numerical evaluation.

\begin{figure*}[t]
\centerline{\includegraphics[width=7.5cm]{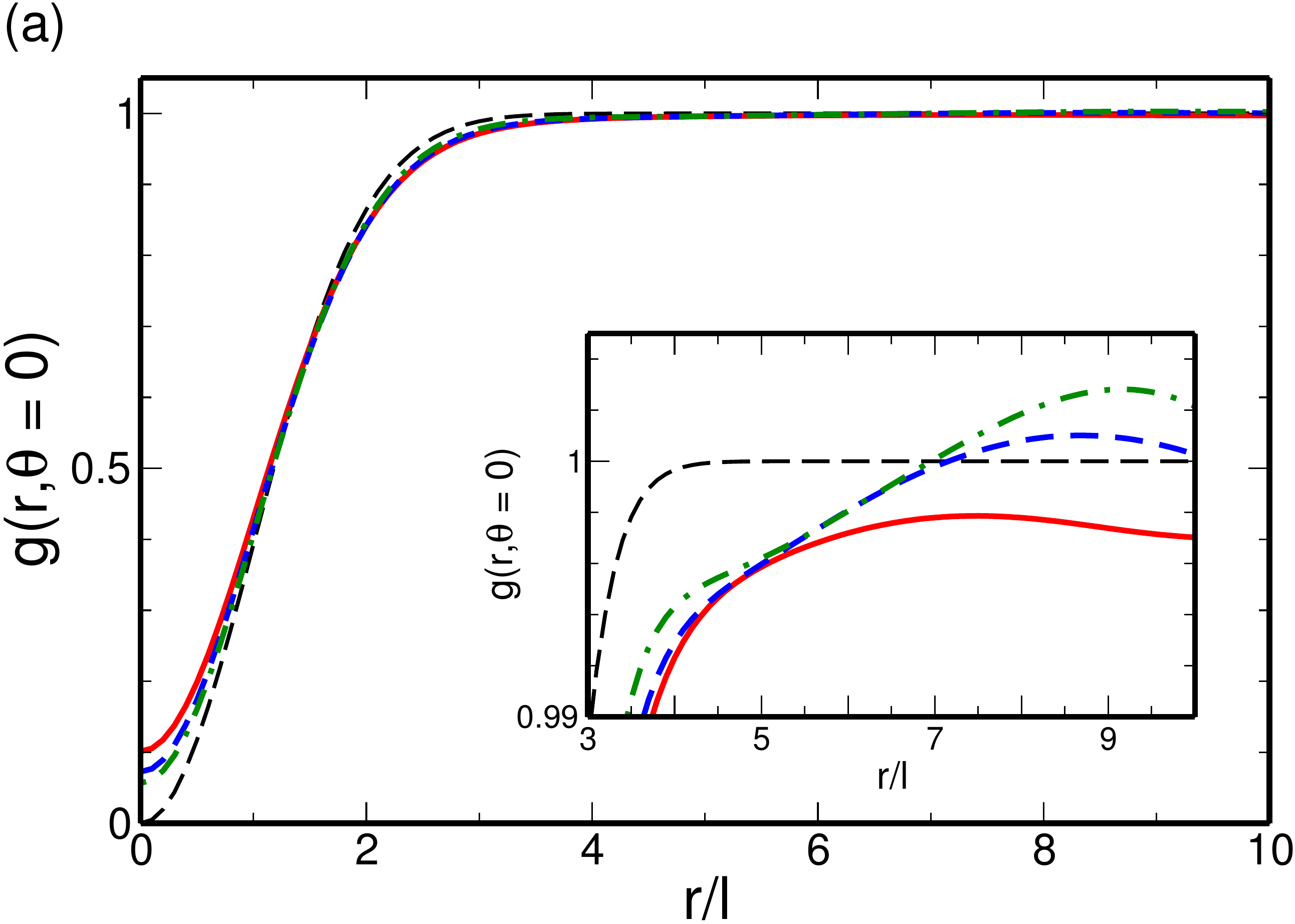}
            \hskip0.7cm
            \includegraphics[width=7.5cm]{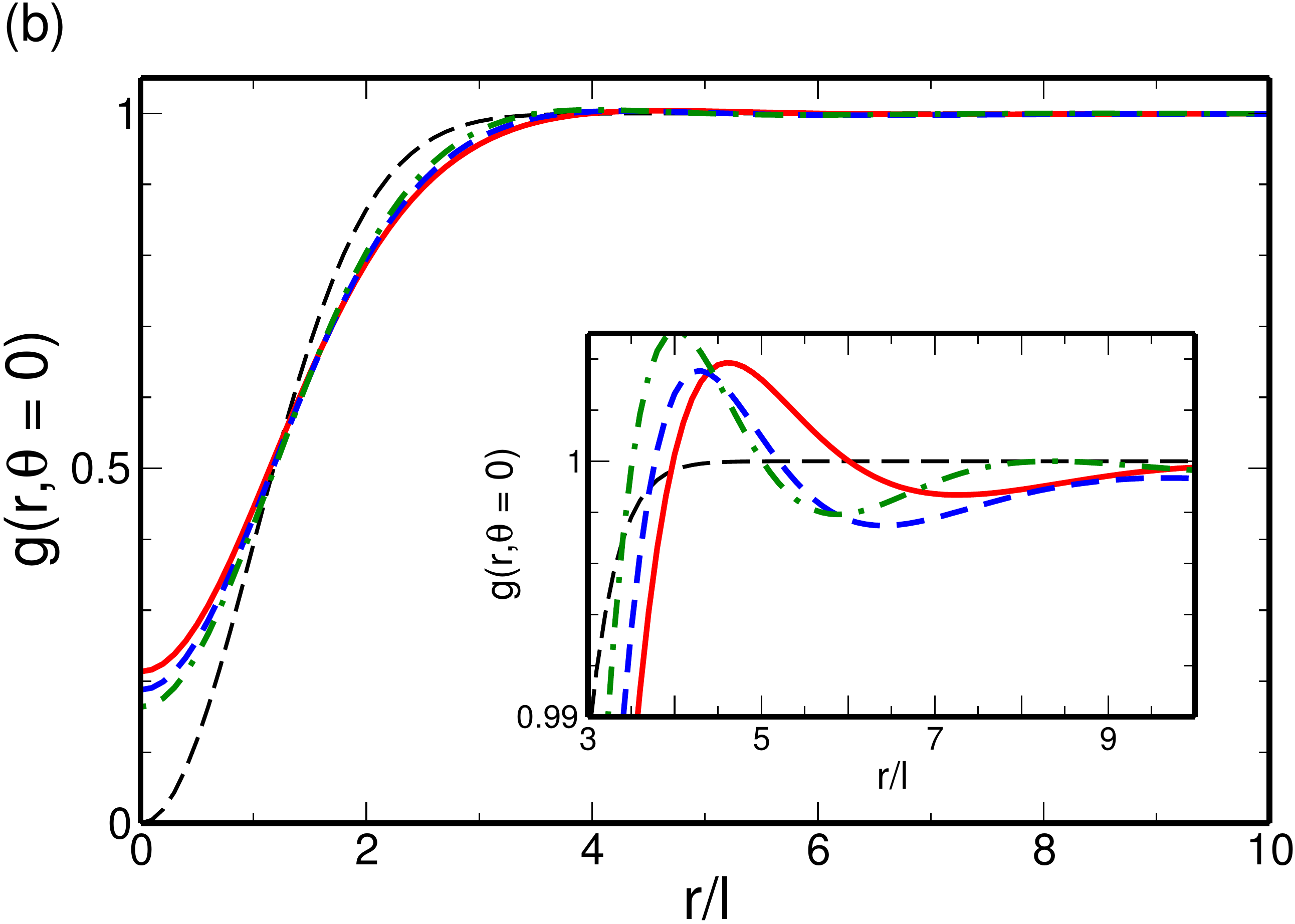}
}
\caption{Pair correlation function \eqref{eq-pairfunction} along one
  particular $\br$--direction. Dashed black line: zero--momentum BEC,
  one--mode approximation with $\mu_0 = 0$,
  Eq.~\eqref{eq-pair-single}. Finite--momentum BEC with 
  (a) $\ell Q = 1.0$ and (b) $\ell Q = 2.0$,
  two--mode approximation. 
  $d = 1$ (solide red line), $1.5$ (long dashed blue line), and $2\,\ell$
  (dot--dashed green line). Insets: details of the corresponding main plots showing the
  behavior of the pair correlation function in the large $r/\ell$ region.}
\label{fig:pair}
\end{figure*}

In Fig.~\ref{fig:pair}, we plot the pair correlation function along
one particular $\br$--direction for the zero--momentum BEC (one--mode
approximation with $\mu_0 = 0$) and for the finite--momentum BECs with
$\ell Q = 1$ and  $2$ at $d = 1.0$, $1.5$, and
$2.0\,\ell$, calculated in the two--mode approximation.
Within the approximations considered here, $g(\br)$ is $d/\ell$
independent for the zero--momentum BEC, see dashed line in
Fig.~\ref{fig:pair}. In this case, the pair
correlation function vanishes as $r \rightarrow 0$, indicating the
existence of a correlation hole around the electron, and it is constant at large
$r$, the same features displayed by the single--layer QHS at $\nu =
1$.\cite{rezayi94} A distinct behavior is found for the
finite--momentum BEC. Note that now $g(0) \not= 0$, indicating that
two electrons (with different pseudospins quantum numbers) 
can be very close to each other, corroborating somehow the
schematic picture for a finite--momentum BEC depicted in
Fig.~\ref{fig:bilayer}(c). Moreover, small oscillations at large
$r/\ell$ are observed, which are characteristic of a composite fermion
Fermi--liquid.\cite{rezayi94} These findings support the proposal that
the ground state of the bilayer QHS in the intermediate $d/\ell$ region
can be described by a finite--momentum BEC of bosons $b$.

As a final remark we would like to mention that in Ref.~\onlinecite{kohn70} 
it is shown that an exciton condensate has diagonal long--range
order. Interestingly, the average value of the density operator is
constant and only the density--density correlation function has
Fourier components of the type $\exp[-i{\bf K}\cdot(\br_1 - \br_2)]$.  
Therefore, based only on Eq.~\eqref{exp-rho}, we would expect that the
finite--momentum BEC of magnetic excitons corresponds to an
inhomogeneous phase. 
However, since the pair correlation function
\eqref{eq-pairfunction}, which is related to the density--density
correlation function \eqref{eq-pair-s}, displays a
behavior characteristic of a liquid, 
we then conclude that the finite--momentum BEC of magnetic excitons is
indeed a homogeneous phase.
The disagreement between our results and the general analysis of
Ref.~\onlinecite{kohn70} might be related to the fact that here the
electrons are restrict to the lowest Landau level. Recall that such
a restriction, e.g.,  modifies the commutation relations between the 
electron and pseudospin density operators.\cite{gmp}

\section{Discussion}
\label{sec:discussion}

\subsection{Relation to previous work}
\label{sec:previous-work}

In this section, we briefly summarize some previous results about the
bilayer QHS and compare them with the ones derived here using the
one and two--mode approximations.

Park\cite{park04} proposed that the bilayer QHS at $\nu_T=1$ develops
a pseudospin spiral long range order at intermediate $d/\ell$ values.
In this case, the main interlayer correlations are established between
electrons and holes localized in different guiding centers [see
Eq.~(8), Ref.~\onlinecite{park04}]. Interestingly, the excitation
spectrum is gapped. 
It is argued that there is no fundamental reason for a Goldstone mode
in this case (see note 14 in Ref.~\onlinecite{park04}).
These two aspects above discussed suggest that the pseudospin
spiral state bears some similarities with the finite--momentum BEC of
bosons $b$. Moreover, it is also conjectured\cite{park04} that the ground
state is indeed given by a bound state between two pseudospin spirals
with opposite winding direction. Recall that by splitting the
finite--momentum condensate into two equal pieces, the total energy of
the system decreases, see Fig.~\ref{fig:egs}.

The ground state energy of finite size systems was calculated within
the exact diagonalization technique.\cite{schliemann03} It is shown
that, regardless the size of the system, the ground state energy is
almost constant for large $d$, a signature of the decoupling
between the two layers. As we can see in Fig.~\ref{fig:egs}, the
ground state energy of a finite--momentum BEC slowly varies for larger
$d/\ell$. Moreover, the variation decreases when the condensate is
separated into two equal pieces.

Nomura and Yoshioka also consider finite size systems and calculate the
pair correlation function \eqref{eq-pairfunction} via exact
diagonalization.\cite{nomura02} It is found that for $d = 0.3\,\ell$,
both $g_{\uparrow\,\uparrow}(\br)$ and $g_{\uparrow\,\downarrow}(\br)$
vanish as $r \rightarrow 0$ but, for a larger $d = 0.9\,\ell$, while
$g_{\uparrow\,\uparrow}(0)$ vanishes, $g_{\uparrow\,\downarrow}(0)$ is
now finite. Concerning the large $r$ region, both
$g_{\uparrow\,\uparrow}(\br)$ and $g_{\uparrow\,\downarrow}(\br)$
seems to be constant for $d = 0.3\,\ell$ but, for $d = 0.9\,\ell$, they
show small oscillations. Note that the pair correlation function,
Fig.~\ref{fig:pair}, qualitatively displays the same features.

Based on a Chern-Simons gauge theory, Bonesteel {\it et
al.}\cite{bonesteel96} show that by approaching two composite fermion
Fermi seas, there is always an instability towards the formation of
composite fermion Cooper pairs. The theory is valid only in the large
$d$ region. The possibility of interlayer composite fermion pairing is
considered in Ref.~\onlinecite{kim01} where some trial wave functions are
discussed. Assuming a $p_x - ip_y$ pairing instability, it is shown
that the two possible wave functions correspond to the $(3,3,-1)$ and
the so--called ``strong'' pairing (SP) states. The former phase displays a
gapped (neutral) excitation spectrum. A qualitative phase diagram is
also proposed and one of the possibilities is that the ground state
changes as $d/\ell$ increases according to the following sequence:
$111$ -- SP -- $(3,3,-1)$ state. 
Unfortunately, it is not clear how to compare a finite--momentum BEC 
of magnetic excitons with the SP and $(3,3,-1)$ states.

Further support for pairing between interlayer composite fermions is
provided in Ref.~\onlinecite{moller08}. However, the numerical results
indicate that a $p_x + ip_y$ pairing may occur instead of the $p_x -
ip_y$ considered by Kim {\it et al.}\cite{kim01}  
Mixed Fermi--Bose trial wave functions were then proposed\cite{moller09} to
describe the intermediate $d \sim \ell$ region, where the bosonic
part is given by the 111 state while the fermionic one, by a paired
composite fermion state. Such an approach indeed follows the lines of an
earlier work by Simon and coworkers,\cite{simon03} where mixed wave
functions are considered, but here the fermionic piece is given by two
composite fermion Fermi seas. 
It is shown that\cite{moller09} the mixed
wave functions with paired states provided a better description for the
intermediate $d \sim \ell$ region than the ones which do not include
pairing.  
Again, it is difficult to compare this mixed Fermi--Bose 
wave functions with the finite--momentum BEC discussed here. 
We would like to point out that the description in terms of a
finite--momentum BEC of magnetic excitons involves only bosonic
degrees of freedom.

Finally, studying the bilayer QHS within a Ginzburg--Landau theory,
Ye and Jiang\cite{ye07} suggested that the ground state is given by a
pseudospin density wave for  $d_{c1} < d < d_{c2}$. In this case, the system
undergoes two first--order quantum phase transitions: from the $111$
state to a pseudospin density wave at $d_{c1}$, and from the latter to
two weakly coupled composite fermion Fermi liquids at $d_{c2}$.
The pseudospin density wave phase proposal is based on an earlier
random phase approximation calculation,\cite{fertig89} which 
finds that the (neutral) excitation spectrum has a minimum
(magneto-roton) at finite momentum $\ell q \sim 1.0$ and that the energy of
this mode vanishes at $d = 1.2\,\ell$. This feature indicates
that a phase that spontaneously breaks translational symmetry may be
realized. Such an inhomogeneous phase is studied, for instance, in
Ref.~\onlinecite{cote92} within the Hartree--Fock
approximation. Interestingly, it is shown that the ground 
state energy is almost $d/\ell$ independent, similar to
Fig.~\ref{fig:egs}. However, we should recall that 
the finite--momentum BEC is indeed a homogeneous phase, see
Sec.~\ref{sec:properties}.

\subsection{Consequences for the bilayer QHS at $\nu_T=1$}
\label{sec:consequences}

As discussed in Sec.~\ref{sec:onemode}, the two approximation schemes
used throughout this paper, the one and two--mode approximations,
impose some strong restrictions on the description of the
finite--momentum BEC of magnetic excitons, but they allow
us to carry out detailed calculations in order to verify the stability
of such a phase. Therefore, they should be seen as an initial approach
to study the finite--momentum BEC. A more elaborated approximation
scheme is needed. 
Since the two--mode approximation suggests that the
system reduces its energy by splitting the condensate into two equal
pieces, a better approximation for the ground state should be 
a finite--momentum condensate such that all modes $\pm\bqz_i$ with $Q_i=Q$ 
and $i = 1, \ldots n$ are (equally) macroscopically occupied, see,
e.g., Refs.~\onlinecite{yukalov78-80} for the case $n\rightarrow \infty$. 
In particular, if $n \rightarrow \infty$, the cylindrical
symmetry of the quasiparticle dispersion relation $\bar{\Omega}_\bq$
would be restored. The implementation of such a scheme is rather
involved and it will be deferred to a future publication.

However, the results that we have derived so far allow us to 
make the following statements about the bilayer QHS: (a) There are
strong indications that a finite--momentum BEC phase is the most stable in the
intermediate $d/l$ region. Such a state bears a strong similarity
with the pseudospin spiral phase proposed by Park.\cite{park04} (b) The
instability of the zero--momentum BEC at $d \sim 0.5\,\ell$,
which we arrive at in Ref.~\onlinecite{doretto06}, indeed corresponds
to a first--order quantum phase transition from a zero--momentum BEC
to a finite--momentum one. In principle, such a transition could be
experimentally observed.

It is also worth mentioning that (c) according to
the two--mode approximation, a finite--momentum BEC with $\ell Q = 1$
is the lowest energy configuration only for $d < 1.6\,\ell$, a value
quite close to the critical $d_c$ where the incompressible--compressible
quantum phase transition is experimentally observed. Curiously, the
Fermi momentum $\bk_F$ of a composite fermion Fermi--liquid at
$\nu=1/2$ is $\ell k_F = 1$.\cite{halperin93} The fact that $\ell Q >
\ell k_F$ for $d > 1.6\,\ell$ could be an indication of the 
incompressible--compressible phase transition.  
However, we should emphasize that this is just an interesting
observation since, at the moment, it is not possible to identify a
composite fermion Fermi--liquid phase within our bosonization formalism.

\subsection{Open questions and next steps}
\label{sec:open}

There are still a couple of open questions that we hope will be
answered once the approximation discussed in the first paragraph of
the previous section is carried out. It remains to be verified 
whether: (i) The features found within the one and two--mode
approximations [points (a) and (b) above] are robust. (ii) The
momentum $\ell Q_0$ associated with the lowest energy configuration
for a given $d/\ell$, Fig.~\ref{fig:qzero}, either varies continuously
with $d/\ell$ or is quantized, $\ell Q_0 = 0, 1, 2, ...$ (iii) It is
possible to identify a finite--momentum BEC with $\ell Q > 1$ with two
composite fermion Fermi--liquids and/or with the pairing states
proposed by M\"oller and coworkers.\cite{moller08} This would be an
important step towards the determination of the nature of the
incompressible--compressible phase transition.

In addition to study a finite--momentum BEC where all modes $\bQ_i$
with $\ell Q_i = \ell Q$ are macroscopically occupied, we also intend
to consider the effects of a (small) finite electron interlayer
tunneling, disorder (hopefully), and the electronic spin. Concerning
the latter, there are some experimental
evidences\cite{exp-spin,giudici08,giudici10} that the  
electron spin degree of freedom might be relevant for a complete
description of the bilayer QHS. For instance, it was recently
reported\cite{giudici08} that the critical $d_c$, where the
incompressible--compressible phase transition takes place, increases
and eventually saturates due to an increasing in-plane magnetic field
${\bf B}_\parallel$. In principle, the effects related to the electronic
spin could be included in our analysis with the help of the
generalized bosonization formalism\cite{doretto07} which has been
recently proposed by two of us to study the quantum Hall effect in
graphene at $\nu=0$ and $\pm 1$.

So far, we have focused on the elementary neutral
excitations of the bilayer QHS. It remains to be checked how {\sl
charged} excitations are described within our bosonization
approach. Such excitations are important when disorder effects are
taken into account (see, e.g., Refs.~\onlinecite{sun10,hyart11} and
the references therein). Two distinct cases should be considered: (a)
Zero--momentum BEC of magnetic excitons. Such a phase can also be seen
as an $XY$ pseudospin ferromagnet.\cite{moon95} In this language,
charged excitations correspond to topological (vortex) excitations
called merons. There are four types of merons: with vorticity $\pm 1$
and electric charge $\pm e/2$. For the single--layer QHS at $\nu=1$,
it was shown\cite{doretto05} that a topological excitation (skyrmion)
can be described as a boson coherent state. Due to the similarities
between the single--layer and the bilayer QHSs, we expect that merons
could be seen as a coherent state of bosons as well. (b)
Finite--momentum BEC of magnetic excitons. Here the mapping into an
$XY$ pseudospin model no longer holds (see note 14 in
Ref.~\onlinecite{park04}) and therefore, it is not yet clear whether
charged excitations could also be described as a boson coherent state.

Finally, concerning experiments, it would be interesting, e.g., to
calculate the interlayer tunneling current for the finite--momentum
BEC phase. The first theoretical works\cite{wen-ezawa93} (clean
limit) indicated that the bilayer QHS should display a Josephson--like
effect, i.e., a zero--bias infinite tunneling conductance should be
observed. Such a feature is related to the gapless linearly dispersing
(neutral) excitation spectrum at low energies associated with the
Halperin $111$ phase (zero--momentum BEC of magnetic excitons).
Instead, an enhanced finite tunneling conductance at zero--bias
voltage was experimentally observed.\cite{spielman00,spielman01} 
In order to account for the experimental features,
disorder effects were then considered. At the moment, the
experimental data have been understood within an $XY$
pseudospin model with the tunneling term \eqref{ham-t} perturbatively
treated and with disorder--induced merons phenomenologically
included in the electron tunneling operator (for a review, see, e.g.,
Ref.~\onlinecite{hyart11} and the references therein).
Interestingly, such a scheme indicates that by adding a parallel
magnetic field ${\bf B}_\parallel$ to the sample, the tunneling
conductance versus bias voltage data could provide a measurement of
the gapless linearly dispersing excitation spectrum. Again,
the experimental data of Spielman {\it et al.},\cite{spielman01} who
found some evidences for the existence of such collective excitation,
were analyzed within the above $XY$ pseudospin framework.

In principle, our results are in contradiction with the experimental
data\cite{spielman01} since we have found a gapped phase in the
intermediate $d/\ell$ region, where tunneling experiments were
performed. However, note that according to our results, a true
Josephson--like effect should occur only at very small $d/\ell$, where
the zero--momentum BEC phase sets in. This is somehow in agreement
with the experiments. Our next task is to verify whether the
gapped excitation spectrum, Figs.~\ref{fig:disp-1mode} and
\ref{fig:disp-2modes}, could account for the observed finite tunneling
conductance at zero bias voltage. In principle, we can calculate the
interlayer tunneling current (clean limit) within our bosonization
formalism, treating the tunneling term \eqref{ham-t} nonpertubatively. 
Disorder effects could be included latter, for instance, following the
lines of Ref.~\onlinecite{sun10}. In this way, we hope we can provide
an alternative interpretation for the experimental data reported in
Refs.~\onlinecite{spielman00,spielman01}.

\section{Summary}
\label{sec:summary}

In this paper, we studied the bilayer QHS at $\nu_T = 1$ within
the bosonization method,\cite{doretto05} a formalism which allows us to
properly treat the magnetic exciton as a boson, and we showed that the
ground state of the system in the region $d \sim \ell$ (zero interlayer
tunneling case) can be seen as a finite--momentum BEC of magnetic
excitons. Our findings are in agreement with previous results which
suggest that an intermediate phase may show up between the
(incompressible) Halperin 111 state (ground state for small $d/\ell$) 
and the (compressible) composite fermion Fermi--liquids (ground state
for larger $d/\ell$).

The stability of the finite--momentum BEC has been analyzed via two
distinct approximation schemes: the one--mode approximation, where it
is assumed that the bosons macroscopically occupied only one momentum $\bQ =
Q\hat{x}$ with $\ell Q \not= 0$, and the two--mode approximation, where both
$\pm\bQ$ modes with $\bQ = Q\hat{x}$ are macroscopically
occupied. We have found that such a phase can be realized as long as
the excitation spectrum is gapped. 
The comparison between the ground state energy curves in terms of the ratio
$d/\ell$ for configurations with different $\ell Q$ 
as well as the analysis of the quasiparticle excitation spectra
provide strong evidences for a first--order quantum phase transition at
small $d/\ell$, i.e., a transition from a zero--momentum BEC, a phase
that we had already analyzed in Ref.~\onlinecite{doretto06} and that
corresponds to Halperin 111 state, to a finite--momentum BEC. 
We hope that such first--order quantum phase transition can be
experimentally observed in the near future. 

As a final remark, we would like to emphasize that the bosonization
method introduced in Ref.~\onlinecite{doretto05} can
be used to study both the single and double--layer QHSs at $\nu=1$. 
In other words, we can describe both systems using the same degree
of freedom, the magnetic exciton, in the limit where this object can
be treated as a boson.


\acknowledgments

We thank M. Vojta, Lars Fritz, and Achilleas Lazarides for helpful
discussions.  
R.L.D. kindly acknowledges FAPESP (project No. 10/00479-6) for the
financial support,  
A.O.C., the partial financial support from CNPq (project
No. 303073/2010-1) and FAPESP (project  No. 07/57630-5), and
C.M.S., the Netherlands Organization for Scientific
Research (NWO).

\appendix

\section{About the bosonization scheme}
\label{ap:boso}

In Ref.~\onlinecite{doretto05}, it is shown that a bosonization
formalism for the two--dimensional electron gas under a strong
magnetic field (single--layer QHS at $\nu=1$) can be developed
following the lines of the bosonization scheme used to describe
one--dimensional electronic systems. It is interesting that this 
formalism also gives quite reasonable results for the bilayer QHS at
$\nu_T=1$ even though such a system is in a limit very far from the
one considered in Ref.~\onlinecite{doretto05}. In this section, we
start providing some heuristic arguments which tell us why the
bosonization scheme\cite{doretto05} can also be employed to study the
bilayer QHS at $\nu_T=1$ and later, we show a simple calculation which
corroborates such arguments.

The bosonization method for the single--layer QHS at $\nu=1$ is based
on the following points: The ground state of the (noninteracting)
two--dimensional electron gas at $\nu=1$ is the quantum Hall
ferromagnet $|{\rm FM}\rangle$, Fig.~\ref{fig:qhfm} (a), the
reference state. Elementary (neutral) excitations are
electron--hole pairs or spin flips, Fig.~\ref{fig:qhfm} (b), which can
be obtained by applying the spin density operator $S^-$
into $|{\rm FM}\rangle$, i.e., $|\Psi\rangle \propto S^-|{\rm FM}\rangle$. 
Although the commutation relation between the spin density operators
$S^+$ and $S^-$ (projected into the lowest Landau level),   
\begin{equation}
[S^+_\bq,S^-_\bk] = e^{l^2qk^*/2}\rho_\uparrow(\bq+\bk) - 
                   e^{l^2kq^*/2}\rho_\downarrow(\bq+\bk)
\label{eq-commutator}
\end{equation}
with $q = q_x + iq_y$, differs from the canonical commutation relation
between a creation and an annihilation boson operators, it is still
possible to define boson operators $b$ as in Eq.~\eqref{boson-op} if we 
follow the lines of Tomonaga's procedure\cite{mahan} for
one--dimensional electron systems. Using the fact that 
\begin{eqnarray}
 \rho_\uparrow(\bq) &\approx& 
        \langle {\rm FM} | \rho_\uparrow(\bq) | {\rm FM} \rangle =
                    N_\phi\delta_{\bq,0},
\nonumber \\
&& \\
 \rho_\downarrow(\bq) &\approx& 
        \langle {\rm FM} | \rho_\downarrow(\bq) | {\rm FM} \rangle =
                            0,
\nonumber
\end{eqnarray}
we realize that Eq.~\eqref{eq-commutator} assumes the form
\begin{equation}
  [S^+_\bq,S^-_\bk] \approx N_\phi\exp[-(lq)^2/2]\delta_{\bq+\bk,0}, 
\label{eq-cummutator2}
\end{equation}
which now resembles a canonical commutation relation for boson
operators. In other words, as long as we are close to
the $|{\rm FM} \rangle$ state, i.e., the number of bosons in the
system is small, electron--hole excitations (magnetic excitons) can be
approximately treated as bosons.

Turning to the bilayer QHS at $\nu_T=1$, we note that such a system is
very far from the limit discussed above because the configuration $\nu_T =
\nu_\uparrow + \nu_\downarrow = 1/2 + 1/2 = 1$ corresponds to a system
with $N_\phi/2$ bosons, see Fig.~\ref{fig:bilayer} (b). As we will see
below, this is indeed a very special configuration where density
fluctuations guarantee that the relation \eqref{eq-cummutator2} still
holds.

Let us consider as a reference state the zero--momentum BEC of
magnetic excitons illustrated in Fig.~\ref{fig:bilayer} (b) and
identify the electronic spin degree of freedom of the single--layer
with the pseudospin one of the bilayer QHS. In this case, we have
\begin{eqnarray}
 \rho_\uparrow(\bq) &\approx& 
        \langle \rho_\uparrow(\bq)  \rangle  + \delta\rho_\uparrow(\bq),
\nonumber \\
&& \label{av-rho}\\
 \rho_\downarrow(\bq) &\approx& 
        \langle \rho_\downarrow(\bq) \rangle + \delta\rho_\downarrow(\bq).
\nonumber
\end{eqnarray}
Since $\langle \rho_\uparrow(\bq) \rangle = \langle \rho_\downarrow(\bq)
\rangle = (N_\phi/2)\delta_{\bq,0}$, we note that here the equivalent of
Eq.~\eqref{eq-cummutator2} vanishes, and therefore the bosons $b$
are no longer well defined. However, since the total filling
factor is fixed, we have
\begin{equation}
  \delta\rho_\uparrow(\bq) = -\delta\rho_\downarrow(\bq) =
  \delta\rho(\bq), 
\label{delta-rho01}
\end{equation}
which indicates that a finite (approximate)
commutation relation can still be obtained if we now consider
density fluctuations in Eq.~\eqref{av-rho}.
Due to the relation between the magnetic exciton momentum
$\bq$ and the guiding centers associate with the electron and the
hole, we have
\begin{equation}
  \delta\rho(\bq) = \delta\rho_{\rm local}(\bq) 
                   + \delta\rho_{\rm nonlocal}(\bq),  
\label{delta-rho02}
\end{equation}
where $\delta\rho_{\rm local}(\bq)$ and $\delta\rho_{\rm
nonlocal}(\bq)$ correspond to density fluctuations on the same and
different guiding centers respectively. Moreover, we can also write 
\begin{equation}
   \delta\rho_{\rm local}(\bq) \approx \rho_0\delta_{\bq,0},
\label{delta-rho-local}
\end{equation}
with $\rho_0$ being a constant, which implies that the relation
\eqref{eq-cummutator2} still holds for the bilayer QHS at $\nu_T=1$.

In order to see that \eqref{delta-rho02} and \eqref{delta-rho-local}
are indeed quite reasonable assumptions, let us calculate the density
fluctuations  
\begin{equation}
 \delta\rho_\alpha(\bq) = \sqrt{ \langle \rho^2_\alpha(\bq) \rangle
                        - \langle \rho_\alpha(\bq) \rangle^2 }
\label{def-delta-rho}
\end{equation}
within the two--mode approximation. It can be seen as a kind of
self-consistent check of the above assumptions.

In the Bogoliubov approximation, the Fourier transform of the
$\alpha$--electron density operator reads 
[see Eqs.~(25) and (26) from Ref.~\onlinecite{doretto05}]  
\begin{eqnarray}
\rho_\alpha(\bq)  &=&  \delta_{\alpha,\uparrow}\delta_{q,0}N_\Phi 
  -\alpha e^{-(\ell q)^2/4} 
   \left[e^{- i\alpha\bq\wedge\bQ/2}\sqrt{N_0}b_{\bQ+\bq}^\dagger
   \right.  
\nonumber \\
&& \left. \right. \nonumber \\
  &+& \left.
     e^{i\alpha\bq\wedge\bQ/2}\sqrt{\bar{N}_0}b_{-\bQ+\bq}^\dagger
     \right.
\nonumber \\
  &+& \left. \sum_{\bk\not=0,-2\bQ}
     e^{-i\alpha\bq\wedge(\bQ + \bk)/2}b_{\bQ+\bk+\bq}^\dagger b_{\bQ+\bk}\right].
\label{rho-boso-sigma} 
\end{eqnarray}
Since
\begin{eqnarray}
 \langle b_{\bQ+\bq}^\dagger \rangle &=& \delta_{\bq,0}N^{1/2}_0
                   + \delta_{\bq,-2\bQ}\bar{N}^{1/2}_0,
\nonumber \\
 \langle b_{-\bQ+\bq}^\dagger \rangle &=& \delta_{\bq,0}\bar{N}^{1/2}_0
                   + \delta_{\bq,2\bQ}N^{1/2}_0,
\nonumber 
\end{eqnarray}
and
\begin{eqnarray}
 \langle b_{\bQ+\bk+\bq}^\dagger b_{\bQ+\bk} \rangle &=& 
  \delta_{\bq,0}\langle b_{\bQ+\bk}^\dagger b_{\bQ+\bk} \rangle
 +  \delta_{\bq,2\bQ}\langle b_{3\bQ+\bk}^\dagger b_{\bQ+\bk} \rangle
\nonumber \\
 &+&  \delta_{\bq,-2\bQ}\langle b_{-\bQ+\bk}^\dagger b_{\bQ+\bk} \rangle,
\nonumber 
\end{eqnarray}
where the expectation value is taken with respect to the ground state
of the bilayer QHS, i.e., the vacuum for the bosons $a$, see
Eq.~\eqref{def-a-bosons}, it follows that 
\begin{equation}
 \langle \rho_\alpha(\bq) \rangle =
  \delta_{\bq,0}\frac{1}{2}N_\phi  
     - (\delta_{\bq,2\bQ} + \delta_{\bq,-2\bQ})e^{-(\ell\bQ)^2}F_\alpha(\bq).
\label{ave-rho-2mode}
\end{equation}
Here,
\begin{eqnarray}
 F_\alpha(\bq) &\equiv& \sqrt{N_0\bar{N}_0} 
\nonumber \\
 &+& \sum_{\bk\not=0,2\alpha\bQ} e^{i\alpha\bq\wedge\bk/2} 
       \left[ v_1(\bk)v_2(\bk) + v_3(\bk)v_4(\bk) \right],
\nonumber
\end{eqnarray}
with $v_i(\bq)$ being the Bogoliubov coefficients, Eq.~\eqref{coef-bogo2}. 
Concerning $\langle \rho^2_\alpha(\bq) \rangle$, after some algebra,
it is possible to show that
\begin{eqnarray}
 \langle \rho_\alpha(\bq)\rho_\alpha(\bq) \rangle &=&
 \delta_{\bq,0}\left[ \frac{1}{4}N^2_\phi + \rho_0 \right]
  + \left[ \delta_{\bq,2\bQ} + \delta_{\bq,-2\bQ} \right]
\nonumber \\
 && \times e^{-2(\ell\bQ)^2}
     \left[ F^2_\alpha(\bq) + \rho_1(\bq) \right],
\label{ave-rho-rho-2mode}
\end{eqnarray}
where
\begin{eqnarray}
 \rho_0 &=& \sum_\bp 
         \left[ v_1^2(\bp) + v_3^2(\bp) \right]^2
       + \left[ u_1(\bp)v_2(\bp) \right. 
\nonumber \\
&& \left. + u_3(\bp)v_4(\bp) \right]^2
   + 2\left[ v_1(\bp)v_2(\bp) + v_3(\bp)v_4(\bp) \right]^2
\nonumber \\
&& \nonumber \\
&& + 2\left[ u_1(\bp)v_1(\bp) + u_3(\bp)v_3(\bp) \right]^2,
\nonumber \\
&& \nonumber \\
 \rho_1(\bq) &=& 
  \sum_\bp \cos(\bq\wedge\bp)\left[v_1(\bp)v_2(\bp) 
                            + v_3(\bp)v_4(\bp) \right]^2
\nonumber \\
  &+& \left[ u_1(\bp)v_2(\bp) + u_3(\bp)v_4(\bp) \right]^2.
\nonumber
\end{eqnarray}
Therefore, Eqs.~\eqref{delta-rho01}, \eqref{def-delta-rho},
\eqref{ave-rho-2mode}, and \eqref{ave-rho-rho-2mode} yield
\begin{equation}
 \delta\rho^2(\bq) = \delta_{\bq,0}\rho_0 
 + \left[ \delta_{\bq,2\bQ} + \delta_{\bq,-2\bQ} \right]
         e^{-2(\ell\bQ)^2}\rho_1(\bq). 
\end{equation}
Note that the first term of the above equation can be identified with
$\delta\rho_{\rm local}(\bq)$ in \eqref{delta-rho02} while the second
one, with $\delta\rho_{\rm nonlocal}(\bq)$. Moreover, one can easily
see that $\rho_0 > e^{-2(\ell\bQ)^2}\rho_1(\pm\bQ)$, which
corroborates the fact that $\delta\rho_{\rm nonlocal}(\bq)$ can be
neglected with respect to $\delta\rho_{\rm local}(\bq)$.   

In the one--mode approximation, Eqs.~\eqref{ave-rho-2mode} and
\eqref{ave-rho-rho-2mode} reduce to  
\begin{eqnarray}
\langle \rho_\alpha(\bq) \rangle &=& \delta_{\bq,0}\frac{1}{2}N_\phi,
\nonumber \\
\langle \rho_\alpha(\bq)\rho_\sigma(\bq) \rangle &=&
 \delta_{\bq,0}\left[ \frac{1}{4}N^2_\phi 
       + \frac{1}{2}\sum_{\bk\not=\bQ}\frac{\epsilon_\bk(\epsilon_\bk
         - \Omega_\bk)}{\Omega^2_\bk} \right].
\nonumber
\end{eqnarray} 
Although $\delta\rho_{\rm nonlocal}(\bq) = 0$, the bosonization
scheme\cite{doretto05} still holds because the relevant term
$\delta\rho_{\rm local}(\bq)$ is finite within this approximation.

\section{Alternative effective boson models for the bilayer QHS}
\label{ap:alternative}

In this section, we briefly comment on some different effective boson
models proposed to describe the bilayer QHS.

In Ref.~\onlinecite{burkov02}, the effects of the electron spin degree
of freedom are taken into account. Here the original fermion model is
mapped into an effective lattice spin--pseudospin model, which is then
analyzed  within a generalized Schwinger boson mean--field
theory. Finite temperature properties are calculated, for instance,
the temperature dependence of the spin and in--plane pseudospin
magnetizations. A proper comparison between our results and the ones
of Ref.~\onlinecite{burkov02} will be possible only after including the
electronic spin in our formalism, see Sec.~\ref{sec:open}. We would
like to recall that our approach is based on a direct mapping of the
interacting fermion model \eqref{ham} into the boson model
\eqref{ham-boso}, no lattice degrees of freedom are introduced.  

Tieleman and co-workers\cite{tieleman09} derived an effective boson
model for the bilayer QHS also following the ideas of the bosonization 
scheme,\cite{doretto05} but with some important differences: The
single particle electron states considered are no longer the
pseudospin up and down lowest Landau levels, see Fig.~\ref{fig:bilayer}, 
but symmetric and antisymmetric linear combination of these
states. Instead of defining the boson operators with respect to the
quantum Hall ferromagnet $|{\rm FM}\rangle$, see Sec.~\ref{sec:boso},
the reference state is the completely filled symmetric
state $|{\rm SYM}\rangle$. Therefore, the boson operators
introduced in Ref.~\onlinecite{tieleman09}, hereafter called $\beta$,
differ from the bosons $b$ discussed in Sec.~\ref{sec:boso}. Most
importantly, the ground state corresponds to an almost filled
symmetric state instead of the BEC of magnetic excitons considered
within our formalism. Since the procedure adopted in
Ref.~\onlinecite{tieleman09} to derive an effective boson model is
not fully consistent with the bosonization scheme,\cite{doretto05} 
below we briefly revisit its derivation.

Formally, the bosonization schemes of Refs.~\onlinecite{doretto05} and
\onlinecite{tieleman09} are identical. Therefore, the bosonic
representations of the electron density and pseudospin density
operators are given by Eqs.~(27)--(29), and (31) of
Ref.~\onlinecite{doretto05} with the replacement $b \rightarrow
\beta$. Substituting these expressions in Eq.~(7) of
Ref.~\onlinecite{tieleman09}, we arrive at  
\begin{equation}
H_B^{SAS} = H_2 + H_4 + H_6,
\label{htotal-sas}
\end{equation} 
where $H_{2,4,6}$ respectively correspond to terms with 2, 4, and 6
boson operators $\beta$. In particular, 
\begin{equation}
H_2 = \sum_\bq \tilde{\epsilon}_\bq \beta^\dagger_\bq \beta_\bq 
         + \frac{1}{2}\left(
         \tilde{\lambda}_\bq \beta^\dagger_\bq \beta^\dagger_{-\bq} 
         + \bar{\lambda}_\bq \beta_{-\bq} \beta_\bq  \right),
\label{h2-sas}
\end{equation}
with
\begin{eqnarray}
\tilde{\epsilon}_\bq &=& t + \tilde{\lambda}_\bq 
                           - \sum_\bk e^{-(lk)^2/2}v_c(k)
\nonumber \\
&& \nonumber \\
              &&  + \sum_\bk 2v_0(k) e^{-(lk)^2/2} \sin^2(\bk\wedge\bq/2),
\nonumber \\
&& \nonumber \\
\tilde{\lambda}_\bq &=&  N_\phi e^{-(\ell p)^2/2}v_c(q),
\label{coef-sas}\\
&& \nonumber \\
\bar{\lambda}_\bq  &=& \tilde{\lambda}_\bq -  \sum_\bk v_c(k)
                       e^{-(lk)^2/2} \cos(\bk\wedge\bq). 
\nonumber 
\end{eqnarray}
Here $t$ and $v_{0/c}(k)$ respectively correspond to $\Delta_{\rm  SAS}$ 
and $v_{0/z}(k)$, see Sec.~\ref{sec:model}. Comparing
Eqs.~\eqref{htotal-sas}--\eqref{coef-sas} with equations (19)--(21)
of Ref.~\onlinecite{tieleman09}, one can see that the former have
extra terms. This is related to the fact that Tieleman {\it et al.}
approximated the bosonic representation of the operator  $S_x(\bk)$ by
linear terms, i.e., $S_x(\bk) \sim \beta^\dagger_\bk + \beta_{-\bk}$,
while here the complete bosonic representation, which in addition
includes a cubic term in boson operators $\beta$, is considered.  We
should mention that the presence of such a cubic term in the bosonic
representation of $S_x(\bk)$ is important because it guarantees that
the bosonic representation of the electron density and spin density
operators obey the correct commutation relations, the so--called
lowest Landau level algebra, see Sec.~II.D of
Ref.~\onlinecite{doretto05} for details. In order to perform a proper
map of the original fermion model into the boson one, we should
consider the complete bosonic expression of the electron density and
spin density operators. After that, approximations can be employed.

Therefore, if we define boson operators with respect to the
state $|{\rm SYM}\rangle$ and try to be consistent with the
bosonization formalism,\cite{doretto05} we then obtain a
non--Hermitian effective boson model to describe the bilayer
QHS. Recall that the procedure adopted in this paper yields a
Hermitian boson model, Eq.~\eqref{ham-boso}.

Although the Hamiltonian \eqref{htotal-sas} is non--Hermitian, 
let us for the moment assume that this is only an artefact of the 
bosonization scheme and determine the spectrum of the elementary
excitations $\Omega_\bq$ within the harmonic approximation, i.e.,
$H_B^{SAS} \approx H_2$. It is possible to diagonalize \eqref{h2-sas}
as done, for instance, in Ref.~\onlinecite{wu59}. One finds
\[
\Omega_\bq = \sqrt{ \tilde{\epsilon}^2_\bq 
                  - \tilde{\lambda}_\bq\bar{\lambda}_\bq }.
\]
However, $\Omega_\bq$ is not well--defined for small momenta: it is
easy to see that 
$\tilde{\epsilon}^2_\bq - \tilde{\lambda}_\bq\bar{\lambda}_\bq < 0$ 
for $q \rightarrow 0$. One concludes that the restriction of
$H_B^{SAS}$ to the quadratic term \eqref{h2-sas} is not a good
approximation. A proper analysis of $H_B^{SAS}$ should take into
account the $H_4$ and $H_6$ terms. Finally, one should mention that
the approach of Ref.~\onlinecite{tieleman09} is more suitable for 
larger $\Delta_{\rm SAS}$ values while our formalism, for the opposite
limit.   

Interestingly, apart from the coefficient of the 
$\beta^\dagger_\bq \beta^\dagger_{-\bq} $
term, the Hamiltonian \eqref{htotal-sas} corresponds to the effective
boson model proposed in Ref.~\onlinecite{macdonald90}: Here
the energy of the elementary excitations is given by 
$\Omega_\bq = (\tilde{\epsilon}^2_\bq - \bar{\lambda}^2_\bq)^{1/2}$,
which agrees with the diagrammatic calculations of Fertig.\cite{fertig89}
Recall that, in this case, the spectrum has a magnetoroton minimum
which vanishes for $d \approx 1.2\,\ell$ ($t=0$ case).

\section{Details: two--mode approximation}
\label{ap:details}

In this appendix, we provide the full expressions of some quantities
which appear in Sec.~\ref{sec:twomodes}. 

After the substitution \eqref{bogo-subs}, Eq.~\eqref{ham-k} acquires
the form  
\begin{eqnarray}
 K &=& K_0 + \frac{1}{4}\sideset{}{'}\sum_\bq \left[
                \epsilon^+_\bq ( b^\dagger_{\bqz + \bq}b_{\bqz +\bq}
                +  b_{-\bqz - \bq}b^\dagger_{-\bqz - \bq} ) \right.
\nonumber \\
&& \nonumber \\
 &+& \left. \epsilon^-_\bq ( b_{\bqz - \bq}b^\dagger_{\bqz -\bq}
       +  b^\dagger_{-\bqz + \bq}b_{-\bqz + \bq} )       \right.
\nonumber \\
&& \nonumber \\
 &+& \left. \gamma_\bq ( b^\dagger_{\bqz + \bq}b_{-\bqz + \bq} + {\rm h.c.}
        + b^\dagger_{-\bqz - \bq}b_{\bqz - \bq} + {\rm h.c.} )    \right.
\nonumber \\
&& \nonumber \\
        &+&  \left. \lambda_\bq ( b^\dagger_{\bqz + \bq}
                        b^\dagger_{\bqz - \bq} + {\rm h.c.} ) \right.
\nonumber \\
&& \nonumber \\
         &+& \left. \bar{\lambda}_\bq (b^\dagger_{-\bqz +
                   \bq}b^\dagger_{-\bqz - \bq} + {\rm h.c.} ) \right.
\nonumber \\
&& \nonumber \\
        &+&  \left. \xi_\bq  (
                  b^\dagger_{-\bqz +\bq}b^\dagger_{\bqz - \bq} + {\rm h.c.}
                 + b^\dagger_{\bqz + \bq}b^\dagger_{-\bqz - \bq} +
                 {\rm h.c.} ) \right].
\nonumber \\
\label{ham-k4}
\end{eqnarray}
The restriction on the sum over momenta indicates that the modes
that satisfy the condition $\pm\bqz \pm \bq \not= \pm\bqz$
are not included. Here
\begin{eqnarray}
  K_0 &=& 2N_0\bar{N}_0v_{2\bqz}(\bqz,\bqz)
        + (\omega_{\bqz} - \mu)(N_0 + \bar{N}_0)
        - \frac{1}{2}\sum_\bq\epsilon_\bq
\nonumber \\
&&\nonumber \\
  \gamma_\bq &=& 4\sqrt{N_0\bar{N}_0}\left[ v_\bq(-\bqz,\bqz)
                        + v_{2\bqz}(\bq,\bqz)\right],
\nonumber \\
&&\nonumber \\
  \lambda_\bq &=& 4N_0v_\bq(\bqz,\bqz), \;\;\;\;\;\;\;
  \bar{\lambda}_\bq = 4\bar{N}_0v_\bq(\bqz,\bqz),
\nonumber \\
&&\nonumber \\
  \xi_\bq &=& 4\sqrt{N_0\bar{N}_0}v_\bq(-\bqz,\bqz),
\label{coef-ham-k2}
\end{eqnarray}
and $\epsilon^\pm_\bq$ are defined in Eq.~\eqref{coef-ham-k}. 
Setting $\bar{N}_0 = N_0$ and using the fact that $n_0 = N_0/N_B = 
4\pi l^2N_0$, we can write
\begin{eqnarray}
 \lambda_\bq &=& \frac{e^2}{\epsilon l}(2n_0)\frac{e^{-(lq)^2/2}}{lq}[1 -
                        e^{-(lq)d/\ell}\cos(\bq\wedge\bqz)],
\nonumber \\
&& \nonumber \\
 \xi_\bq &=& \frac{e^2}{\epsilon l}(2n_0)\frac{e^{-(lq)^2/2}}{lq}[
                        \cos(\bq\wedge\bqz) - e^{-(lq)d/\ell}],
\label{coef-ham-k22} \\
&& \nonumber \\
 \gamma_\bq &=& \xi_\bq
             + \frac{e^2}{\epsilon l}n_0\frac{e^{-(lq)^2/2}}{lq}[
                          1 - e^{-2(lq)d/\ell}]\cos(\bq\wedge\bqz),
\nonumber  
\end{eqnarray}
which are useful expressions for the numerical calculations.

Concerning the chemical potential, Eq.~\eqref{eq-mu2}, the terms $\mu_0$
and $\mu_1$ are defined as 
\begin{eqnarray}
 \mu_0 &=& \frac{1}{8N_0} \sum_\bq  \left(\frac{1}{\Omega^+_\bq}
              + \frac{1}{\Omega^-_\bq}\right)
    \left[ \lambda_\bq(\epsilon_\bq - \lambda_\bq) +
                           \gamma^2_\bq - \xi^2_\bq \right]
\nonumber \\
&& \nonumber \\
     && - 2\lambda_\bq
        + \frac{1}{D_\bq}\left(\frac{1}{\Omega^+_\bq} - \frac{1}{\Omega^-_\bq}\right)
          \left[ \Delta^2_\bq\lambda_\bq(\epsilon_\bq - \lambda_\bq) \right.
\nonumber \\
&& \left. \right. \nonumber \\
     &&   \left. + (\gamma_\bq\epsilon_\bq - \gamma_\bq\xi_\bq)(\gamma_\bq\epsilon_\bq
          + \gamma_\bq\lambda_\bq - 2\lambda_\bq\xi_\bq) \right],
\label{mu0-2modes} \\
&& \nonumber \\
 \mu_1 &=& 2N_0v_{2\bqz}(\bqz,\bqz) = 
           \frac{e^2}{\epsilon l}\frac{1}{2}n_0\frac{e^{-2(lQ)^2}}{lQ}[1 -
                        e^{-2(lQ)d/\ell}].
\nonumber
\end{eqnarray}

Finally, following the procedure described in Ref.~\onlinecite{blaizot} 
(see Chap.~3) and after some algebra, it is possible to show that the
Bogoliubov coefficients $u_i(\bq)$ and $v_i(\bq)$, the elements of the
$4\times 4$ matrix $\hat{M}_\bq$, Eq.~\eqref{m-matrix}, are given by
\begin{eqnarray}
  u^2_i(\bq), v^2_i(\bq) &=& \frac{1}{4\Omega^i_\bq} 
               \left( \epsilon_\bq + \alpha_i\Delta_\bq \pm \Omega^i_\bq \right)
\nonumber \\
&& \nonumber \\
 &-& \frac{(-1)^i}{4\Omega^i_\bq D_\bq} 
               \left[\gamma_\bq\left(\gamma_\bq\epsilon_\bq - \lambda_\bq\xi_\bq \right)
     - \alpha_i\Delta_\bq\lambda^2_\bq \right.
\nonumber \\
&& \nonumber \\
  &+& \left. \Delta_\bq\epsilon_\bq\left( \alpha_i\epsilon_\bq + \Delta_\bq
     \pm \alpha_i\Omega^i_\bq \right) \right]
\label{coef-bogo2}
\end{eqnarray}
with $i=1$, $2$, $3$, $4$, $\Omega^i_\bq = \Omega^+_\bq$ ($i=1,
3$), $\Omega^-_\bq$ ($i=2, 4$), and $\alpha_i = 1$ ($i=1, 2$),  
$-1$ ($i=3, 4$).

\section{Pair correlation function}
\label{ap:pair}

In this section, we present the full expressions for the pair
correlation function \eqref{eq-pairfunction}, calculated both at the
one and two--mode approximations.

In order to calculate the static structure factor \eqref{eq-pair-s},
it is necessary to project the product $\rho(-\bq)\rho(\bq)$ into the
lowest Landau level. We have [see Eqs.~(4.17) and (4.18) from 
Ref.~\onlinecite{gmp}]  
\begin{equation}
\overline{\rho(-\bq)\rho(\bq)} = \bar{\rho}(-\bq)\bar{\rho}(\bq)
                          + N[1 - \exp(-|l\bq|^2/2)],
\end{equation}
and therefore Eq.~\eqref{eq-pair-s} assumes the form
\begin{equation}
S(\bq) = \bar{S}(\bq) + [1 - \exp(-|l\bq|^2/2)]
\end{equation}
with $\bar{S}(\bq) = (1/N)\langle\bar{\rho}(-\bq)\bar{\rho}(\bq)
\rangle - N\delta_{\bq,0}$ being the projected static structure
factor. From Eq.~\eqref{rho-boso}, it follows that ($\bq\not= 0$) 
\begin{eqnarray}
 \bar{S}(\bq) &=& \frac{4e^{-|lq|^2/2}}{N_\phi}\sum_{\bp,\bk}
                  \sin(\bq\wedge\bp/2)\sin(\bq\wedge\bk/2)
\nonumber \\
 &&\times \langle b^\dagger_{\bk-\bq}b_\bk b^\dagger_{\bp+\bq}b_\bp
\rangle.  
\end{eqnarray}
Here the $\exp(-|lq|^2/4)$ factor is restored in the bosonic
representation of the electron density operator, see Eq.~(27) from
Ref.~\onlinecite{doretto05}. 

Using Wick's theorem, the expectation value in the above equation can
be easily calculated. Performing the replacement
$b^\dagger_{\bqz},b_{\bqz} \rightarrow \sqrt{N_0}$ as done in
Sec.~\ref{sec:onemode} and keeping only the terms with two boson
operators $b$, it is possible to show that in the one--mode
approximation the projected static structure factor assumes the form 
\begin{equation}
  \bar{S}(\bq) \approx 2n_0e^{-|lq|^2/2}\sin^2(\bq\wedge\bqz/2)(
                 v^2_\bq + u^2_\bq - 2u_\bq v_\bq)
\end{equation}
with the Bogoliubov coefficients $u_\bq$ and $v_\bq$ given by
Eq.~\eqref{coef-bogo1}. Note that $\bar{S}(\bq)$ vanishes if $\bqz =
0$ implying that, within this level of approximation, 
\begin{equation}
  g(r) = 1 - \exp[-r^2/(2l^2)]
\label{eq-pair-single}
\end{equation}
regardless the value of $d/\ell$. Eq.~\eqref{eq-pair-single} is
nothing else but the pair correlation function for the single--layer
QHS at $\nu=1$.  

The same procedure can be used to calculate $g(r)$ in the two--mode
approximation. After some algebra, we find that
\begin{eqnarray}
   \bar{S}(\bq) &\approx& 2n_0e^{-|lq|^2/2}\sin^2(\bq\wedge\bqz)\left[
                  \right. 
\nonumber \\
  && \left.   [u_1(\bq) + v_1(\bq) - u_2(\bq) - v_2(\bq)]^2 \right.
\nonumber \\
  &+& \left.   [u_3(\bq) + v_3(\bq) - u_4(\bq) - v_4(\bq)]^2\right], 
\end{eqnarray}
where the Bogoliubov coefficients $u_i(\bq)$ and $v_i(\bq)$ are now
given by Eq.~\eqref{coef-bogo2}.



\begin{thebibliography}{99}

\bibitem{book-qhe} For an introduction, see
{\sl Perspectives in Quantum Hall Effects}, edited by S.\ Das\ Sarma
and A.\ Pinczuk Wiley, New York (1997); {\sl The quantum Hall effect},
D.\ Yoshioka, Springer, Berlin (2002).


\bibitem{eisenstein04} J. P. Eisenstein and A. H. MacDonald, Nature
{\bf 432}, 691 (2004); J. P. Eisenstein, Science {\bf 305}, 950
(2004).


\bibitem{exp-bilayer}
S. Q. Murphy, J. P. Eisenstein, G. S. Boebinger, L. N. Pfeiffer, and
K. W. West, Phys. Rev. Lett. {\bf 72}, 728 (1994);
M. Kellogg, I. B. Spielman, J. P. Eisenstein, L. N. Pfeiffer, and K.
W. West, Phys. Rev. Lett. {\bf 88}, 126804 (2002);
M. Kellogg, J. P. Eisenstein, L. N. Pfeiffer, and K. W. West, Phys.
Rev. Lett. {\bf 93}, 036801 (2004).
Y. Yoon, L. Tiemann, S. Schmult, W. Dietsche, K. von Klitzing, and
W. Wegscheider, 
Phys. Rev. Lett. {\bf 104}, 116802 (2010). 


\bibitem{spielman00}
I. B. Spielman, J. P. Eisenstein, L. N. Pfeiffer, and K. W. West,
Phys. Rev. Lett. {\bf 84}, 5808 (2000). 


\bibitem{spielman01}
I. B. Spielman, J. P. Eisenstein, L. N. Pfeiffer, and K. W. West,
Phys. Rev. Lett. {\bf 87}, 036803 (2001). 


\bibitem{moller09}
G. M\"oller, S. H. Simon, and E. H. Rezayi,
Phys. Rev. B {\bf 79}, 125106 (2009).


\bibitem{schliemann01}
J. Schliemann, S. M. Girvin, and A. H. MacDonald,
Phys. Rev. Lett. {\bf 86}, 1849 (2001).


\bibitem{ye07}
J. Ye and L. Jiang,
Phys. Rev. Lett. {\bf 98}, 236802 (2007).


\bibitem{halperin83}
B. I. Halperin, Helv. Phys. Acta {\bf 56}, 75 (1983).


\bibitem{halperin93}
B. I. Halperin, Patrick A. Lee, and N. Read,
Phys. Rev. B {\bf 47}, 7312 (1993).


\bibitem{heinonen}
O. Heinonen, ed., {\sl Composite Fermions},
World Scientific, Singapore (1998).


\bibitem{fertig89}
H. A. Fertig, Phys. Rev. B {\bf 40}, 1087 (1989).


\bibitem{doretto05}
R. L. Doretto, A. O. Caldeira, and S. M. Girvin,
Phys. Rev. {\bf B} 71, 045339 (2005).


\bibitem{doretto06}
R. L. Doretto, A. O. Caldeira, and C. Morais Smith,
Phys. Rev. Lett. {\bf 97}, 186401 (2006).


\bibitem{simon03}
S. H. Simon, E. H. Rezayi, and M. V. Milovanovic,
Phys. Rev. Lett. {\bf 91}, 046803 (2003).


\bibitem{moller08}
G. M\"oller, S. H. Simon, and E. H. Rezayi, 
Phys. Rev. Lett. {\bf 101}, 176803 (2008).


\bibitem{cote92} 
R. C\^ot\'e, L. Brey, and A. H. MacDonald, 
Phys. Rev. B {\bf 46}, 10239 (1992).


\bibitem{kim01}
Y. B. Kim, C. Nayak, E. Demler, N. Read, and S. Das Sarma,
Phys. Rev. B {\bf 63}, 205315 (2001).



\bibitem{nomura02}
K. Nomura and D. Yoshioka,
Phys. Rev. B {\bf 66}, 153310 (2002).


\bibitem{schliemann03}
J. Schliemann,
Phys. Rev. B {\bf 67}, 035328 (2003).


\bibitem{park04}
K. Park,
Phys. Rev. B {\bf 69}, 045319 (2004).


\bibitem{milica07}
%
%
M. V. Milovanovi\'c,
Phys. Rev. {\bf B} 75, 035314 (2007). 


\bibitem{note01}
The dispersion relation of the free bosons in Ref.~\onlinecite{doretto06}
has an extra term $-(e^2/\epsilon\ell)(d/\ell)$. This contribution is
related to the $\bk=0$ term of the interacting potential
\eqref{ham-i} which was not properly removed in our first
analysis. There are no modifications in the results obtained in
Ref.~\onlinecite{doretto06} since they all depend on the difference
$\omega_\bq - \omega_0$.


\bibitem{kallin84}
C. Kallin and B. I. Halperin,
Phys. Rev. B {\bf 30}, 5655 (1984).


\bibitem{stanescu08}
T. D. Stanescu, B. Anderson, and V. Galitski,
Phys. Rev. A {\bf 78}, 023616 (2008). 


\bibitem{liberto11}
M. Di Liberto, O. Tieleman, V. Branchina, and C. Morais Smith,
Phys. Rev. A {\bf 84}, 013607 (2011). 


\bibitem{fetter}
A. L. Fetter and J. D. Walecka,
{\sl Quantum Theory of Many Particle Systems},
Dover Publications, Mineola, New York (2003).


\bibitem{stoof93}
H. T. C. Stoof and M. Bijlsma,
Phys. Rev. E {\bf 47}, 939 (1993).


\bibitem{shi98}
For a comprehensive review, see H. Shi and A. Griffin,
Phys. Rep. {\bf 304}, 1 (1998).


\bibitem{tinkham}
M. Tinkham, {\sl Introduction to Superconductivy}, 2nd ed.
(McGraw-Hill, New York, 1996).


\bibitem{blaizot}
J. P. Blaizot and G. Ripka, {\sl Quantum Theory of Finite
Systems} (MIT, Cambridge, MA, 1986).


\bibitem{nozieres}
P. Nozi\`eres, in {\sl Bose--Einstein Condensation}, 
edited by A. Griffin, D. W. Snoke, and S. Stringari
(Cambridge University Press, Cambridge, 1996). 



\bibitem{mahan}
G. Mahan, {\sl Many Particle Physics} (Plenum, New York, 2000).


\bibitem{rezayi94}
E. Rezayi and N. Read,
Phys. Rev. Lett. {\bf 72}, 900 (1994). 

\bibitem{kohn70}
W. Kohn and D. Sherrington,
Rev. Mod. Phys. {\bf 42}, 1 (1970).


\bibitem{gmp}
S. M. Girvin, A. H. MacDonald, and P. M. Platzman, 
Phys. Rev. B {\bf 33}, 2481 (1986). 

\bibitem{bonesteel96}
N. E. Bonesteel, I. A. McDonald, and C. Nayak,
Phys. Rev. Lett. {\bf 77}, 3009 (1996).



\bibitem{yukalov78-80}
V. I. Yukalov,
Theor. Math. Phys. {\bf 37}, 1093 (1978);
V.I. Yukalov,
Physica A {\bf 100}, 431 (1980).


\bibitem{exp-spin}
I. B. Spielman, L. A. Tracy, J. P. Eisenstein, L. N. Pfeiffer, and K. 
W. West, 
Phys. Rev. Lett. {\bf 94}, 076803 (2005);
N. Kumada, K. Muraki, K. Hashimoto, and Y. Hirayama, 
Phys. Rev. Lett. 94, 096802 (2005).


\bibitem{giudici08}
P. Giudici, K. Muraki, N. Kumada, Y. Hirayama, and T. Fujisawa
Phys. Rev. Lett. {\bf 100}, 106803 (2008). 


\bibitem{giudici10}
P. Giudici, K. Muraki, N. Kumada, and T. Fujisawa
Phys. Rev. Lett. {\bf 104}, 056802 (2010) 


\bibitem{doretto07}
R. L. Doretto and C. Morais Smith,
Phys. Rev. B {\bf 76}, 195431 (2007).


\bibitem{sun10}
J. Sun, G. Murthy, H. A. Fertig, and N. Bray-Ali,
Phys. Rev. {\bf B} 81, 195314 (2010).

\bibitem{hyart11}
T. Hyart and B. Rosenow,
Phys. Rev. B {\bf 83}, 155315 (2011). 

\bibitem{moon95}
K. Moon, H. Mori, K. Yang, S. M. Girvin, A. H. MacDonald, L. Zheng,
D. Yoshioka, and S.-C. Zhang, 
Phys. Rev. B {\bf 51}, 5138 (1995).

\bibitem{wen-ezawa93}
X. G. Wen and A. Zee, 
Phys. Rev. {\bf B} 47, 2265 (1993);
Z. F. Ezawa and A. Iwazaki,
Phys. Rev. {\bf B} 48, 15189 (1993). 

\bibitem{burkov02}
A. A. Burkov and A. H. MacDonald,
Phys. Rev. B {\bf 66}, 115320 (2002); 
A. Burkov, J. Schliemann, A.H. MacDonald, and S.M. Girvin,
Physica E {\bf 12}, 28 (2002).  

\bibitem{tieleman09}
O. Tieleman, A. Lazarides, D. Makogon, and C. Morais Smith,
Phys. Rev. B {\bf 80}, 205315 (2009).
 
\bibitem{wu59}
T. T. Wu,
Phys. Rev. {\bf 115}, 1390 (1959).

\bibitem{macdonald90}
A. H. MacDonald, P. M. Platzman, and G. S. Boebinger, 
Phys. Rev. Lett. {\bf 65}, 775 (1990). 




\end{thebibliography}
\end{document}